\documentclass[entropy,review,accept,pdftex,moreauthors]{Definitions/mdpi}
\usepackage{comment} 

\usepackage{amsmath}
\usepackage{mathrsfs}
\usepackage{algpseudocodex,algorithm}
\usepackage{changepage}  

\newcommand{\Xb}{{\boldsymbol X}}
\newcommand{\Yb}{{\boldsymbol Y}}

\newcommand{\Rd}{\mathbb{R}^d}
\def\rmdx{\mathrm{d} \boldsymbol{x}}

\def\pxt{{p}(\boldsymbol{x},t)}
\def\bxt{\boldsymbol{b}(\boldsymbol{x},t)}
\def\vxt{\boldsymbol{v}(\boldsymbol{x},t)}
\def\gxt{{g}(\boldsymbol{x},t)}

\def\dxt{\boldsymbol{\sigma}(\boldsymbol{x},t)}

\def\axt{\boldsymbol{a}(\boldsymbol{x},t)}

\def\xt{{\boldsymbol X}_t}

\def\rmd{\mathrm{d}}
\def\xt{{\boldsymbol x}_t}

\firstpage{1} 
\pubvolume{27}
\issuenum{5}
\articlenumber{453}
\pubyear{2025}
\copyrightyear{2025}
\externaleditor{Matteo Convertino} 
\datereceived{14 March 2025} 
\daterevised{18 April 2025} 
\dateaccepted{19 April 2025} 
\datepublished{22 April 2025} 
\hreflink{https://doi.org/10.3390/e27050453} 

\Title{Integrating Dynamical Systems Modeling with Spatiotemporal scRNA-Seq Data Analysis}

\TitleCitation{Integrating Dynamical Systems Modeling with Spatiotemporal scRNA-Seq Data Analysis}

\Author{{Zhenyi} 
 Zhang  $^{1, \dagger}$\orcidA{}, Yuhao Sun $^{2,\dagger}$, Qiangwei Peng $^{1,\dagger}$\orcidC{}, Tiejun Li $^{1,2,3,}$* and Peijie Zhou $^{2,4,5,6,}$*\orcidB{}}

\AuthorNames{Zhenyi Zhang, Yuhao Sun, Qiangwei Peng, Tiejun Li and Peijie Zhou}

\AuthorCitation{Zhang, Z.; Sun, Y.; Peng, Q.; Li, T.; Zhou, P.}

\address{%
$^{1}$ \quad {School} 
 of Mathematical Sciences, Peking University, Beijing 100871, China; \mbox{zhenyizhang@stu.pku.edu.cn (Z.Z.);} qiangwei\_peng@stu.pku.edu.cn (Q.P.)\\
$^{2}$ \quad Center for Machine Learning Research, Peking University, Beijing 100871, China; lh13210817312@gmail.com \\

 $^{3}$ \quad Laboratory of Mathematics and Its Applications (LMAM), Peking University, \mbox{Beijing 100871, China} \\
 $^{4}$ \quad Center for Quantitative Biology, Peking University, Beijing 100871, China \\
$^{5}$ \quad National Engineering Laboratory for Big Data Analysis and Applications, Peking University, \mbox{Beijing 100871, China} \\
$^{6}$ \quad AI for Science {Institute,} 
 Beijing 100080, China}

\corres{Correspondence:  tieli@pku.edu.cn (T.L.); pjzhou@pku.edu.cn (P.Z.)}
\firstnote{These authors contributed equally to this work.}

\abstract{Understanding the dynamic nature of biological systems is fundamental to deciphering cellular behavior, developmental processes, and disease progression. Single-cell RNA sequencing (scRNA-seq) has provided static snapshots of gene expression, offering valuable insights into cellular states at a single time point. Recent advancements in temporally resolved scRNA-seq, spatial transcriptomics (ST), and time-series spatial transcriptomics (temporal-ST) have further revolutionized our ability to study the spatiotemporal dynamics of individual cells. These technologies, when combined with computational frameworks such as Markov chains, stochastic differential equations (SDEs), and generative models like optimal transport and Schrödinger bridges, enable the reconstruction of dynamic cellular trajectories and cell fate decisions. This review discusses how these dynamical system approaches offer new opportunities to model and infer cellular dynamics from a systematic perspective.
}


\begin{document}

\section{Introduction}
Understanding the dynamic change of biological systems has played a central role in life sciences, with~important applications in developmental biology, disease modeling, and~medicine~\cite{lei2023mathematical_review,Xingjianhua_review1,Xingjianhua_review2,schiebinger2021reconstructing,heitz2024spatialreview}. One key framework for understanding these dynamic processes is Waddington’s developmental landscape~\cite{waddington2014strategy,moris2016transition,maclean2018exploring}, which illustrates how cells navigate various potential fates as they differentiate during development. However, how to construct such developmental landscapes or understand the cellular dynamics within the biological systems, presents a significant challenge. To~fully understand these complex cellular transitions, a~deep understanding of gene expression at the single-cell level is essential. Advancements in high-throughput sequencing technologies have enabled unprecedented resolutions into the molecular signatures of individual cells, with~single-cell RNA sequencing (scRNA-seq) emerging as a revolutionary tool~\cite{ziegenhain2017comparative,tang2009mrna,stark2019rna}. scRNA-seq allows for the dissection of cellular heterogeneity and the identification of transcriptional programs underlying complex biological processes, offering a snapshot of gene expression in single cells at a given moment. Despite its powerful capabilities, traditional scRNA-seq provides only a static picture of gene expression across individual cells, missing the temporal information for understanding how cells transition through different~states.

In recent years, the~development of temporally resolved scRNA-seq technologies has begun to gain increasing attention, enabling the capture of gene expression profiles across multiple time points~\cite{nature_review_gene_temporal, schiebinger2021reconstructing, bunne2024optimal}. Another breakthrough in transcriptomics is spatial transcriptomics (ST), which integrates spatial context into gene expression data by mapping RNA profiles within tissue architectures~\cite{staahl2016visualization, rodriques2019slide, stickels2021highly, chen2022spatiotemporal, oliveira2024characterization, moffitt2018molecular, eng2019transcriptome, wang2018three}. When combined with temporal resolution, this approach leads to temporally resolved spatial transcriptomics (temporal-ST), which provides an enhanced tool for studying the spatiotemporal dynamics of single cells~\cite{liu2024spatiotemporal}.

Extracting meaningful dynamical features from spatiotemporal single-cell transcriptomic data remains a significant challenge. Since the inherently destructive nature of single-cell sequencing, each cell can only be measured once during the dynamical process. As~a result, continuous dynamics cannot be directly obtained from the data. Even with temporally resolved single-cell RNA sequencing, we can only obtain unpaired gene expression snapshots at discrete time points, capturing cell distribution changes over time rather than the continuous movement of individual cells. Consequently, inferring cell-state transitions and dynamic regulatory mechanisms from such snapshot-based data necessitates computational modeling approaches, which is an important problem in computational system biology and has gained increasing~importance. 

To address the challenges, numerous computational frameworks have been developed. For~single-cell transcriptomics data, several methods have been proposed to approximate cellular trajectories and dissect dynamic cellular states. Pseudotime inference \mbox{methods~\cite{qiu2017monocle,cao2019singlemonocle3,street2018slingshot},} for~instance, arrange snapshot data along an inferred developmental axis, offering a continuous perspective of cell-state transitions over time. In~addition, RNA velocity analysis~\cite{la2018rna,RNA_velocity_review,wang2024phylovelo,liu2024resolving_review,bergen2020generalizing} has emerged as a powerful tool for understanding cellular dynamics by leveraging splicing kinetics to infer the direction of future gene expression changes. 
Recently, with~the development of temporally resolved sequencing technology, there has been a growing interest in dissecting single-cell dynamics from multiple snapshot data. Simultaneously, the~development of generative modeling techniques, such as diffusion models~\cite{ho2020denoising,sohl2015deep,song2020score,ren2025zeroth}, optimal transport theory~\cite{bunne2024optimal,waddingot,moscot}, flow-based model~\cite{cfm_lipman,cfm_tong}, and~the Schrödinger bridge problem~\cite{sb_Gentil,DeepRUOT} have emerged as key mathematical frameworks for modeling distribution transitions in dynamic biological systems.  More recently, the~rapid development of spatial transcriptomics has opened exciting new avenues for integrating spatial and temporal data. Extending these computational methods to the ST data and capturing spatiotemporal cellular transitions also has inspired many recent kinds of research~\cite{peng2024stvcr}.

Recently, many reviews have provided comprehensive summaries of the methods and advancements in the study of single-cell dynamics. For~example,~{ref.} 
 \cite{wang2021current} reviewed various pseudotime inference methods.~{Ref.}~\cite{saelens2019comparison} conducted a comprehensive benchmarking study on pseudotime inference methods. ~{Refs.}~\cite{wang2021current, li2020mathematics,RNA_velocity_review} reviewed RNA velocity methods in single-cell transcriptomics.~\cite{RNA_velocity_review}  also discussed the limitations and potential extensions of RNA velocity.~{Refs.}~\cite{bunne2024optimal,heitz2024spatialreview,schiebinger2021reconstructing,JiangWANreview} provided an in-depth analysis of the application of optimal transport theory in single-cell or spatial omics data. Additionally,~{refs.}~\cite{Xingjianhua_review2,lei2023mathematical_review} examined various perspectives on cellular dynamics, exploring how the reconstruction of cell states and energy landscapes can contribute to our understanding of cellular behavior and development. The~current review takes a distinct perspective by systematically discussing modeling strategies for different types of data from a dynamical modeling perspective, aiming to unify and expand upon the current methodologies in the~field.

This paper mainly focuses on how dynamic insights can be extracted from high-resolution biological data, including scRNA-seq, temporally resolved scRNA-seq, spatial transcriptomics (ST), and~temporally resolved spatial transcriptomics. We examine how key concepts from dynamical systems modeling—such as Markov chains, stochastic differential equations (SDEs), ordinary differential equations (ODEs), and~partial differential equations (PDEs)—can be effectively applied to the analysis of cellular processes reflected in these high-dimensional data. Furthermore, we explore the application of emerging generative modeling techniques, including optimal transport theory, flow matching, and~the Schrödinger bridge problem, as~approaches for inferring spatiotemporal cellular trajectories and transitions. By~focusing on these modeling strategies, this review aims to provide a systematic framework for understanding cellular dynamics across different types of data, thus advancing the study of spatiotemporal biological~processes.
 
This paper is organized as follows: In Section~\ref{sec:2Overview}, we provide an overview of the data and models, laying the foundation for understanding the types of biological data and the mathematical frameworks. Section~\ref{sec:snaphots} delves into the dynamic modeling of single-cell transcriptomics, with~a focus on both single-cell RNA sequencing (scRNA-seq) and temporal-scRNA-seq. In~Section~\ref{sec:st}, we explore the dynamic modeling of spatial transcriptomics, examining both snapshot-based and temporally resolved approaches to analyze the spatial and temporal dynamics of gene expression. Section~\ref{sec:extensions} discusses the extensions, challenges, and~future directions in the field, highlighting the key limitations and opportunities for advancing the study of cellular dynamics. Finally, we summarize the insights and outline potential areas for future research in Section~\ref{sec:discuss}.

\section{Overview of the Data and~Models}\label{sec:2Overview}
\begin{figure}[H]

\begin{adjustwidth}{-\extralength}{0cm}
\centering 
        \includegraphics[width=0.9\linewidth]{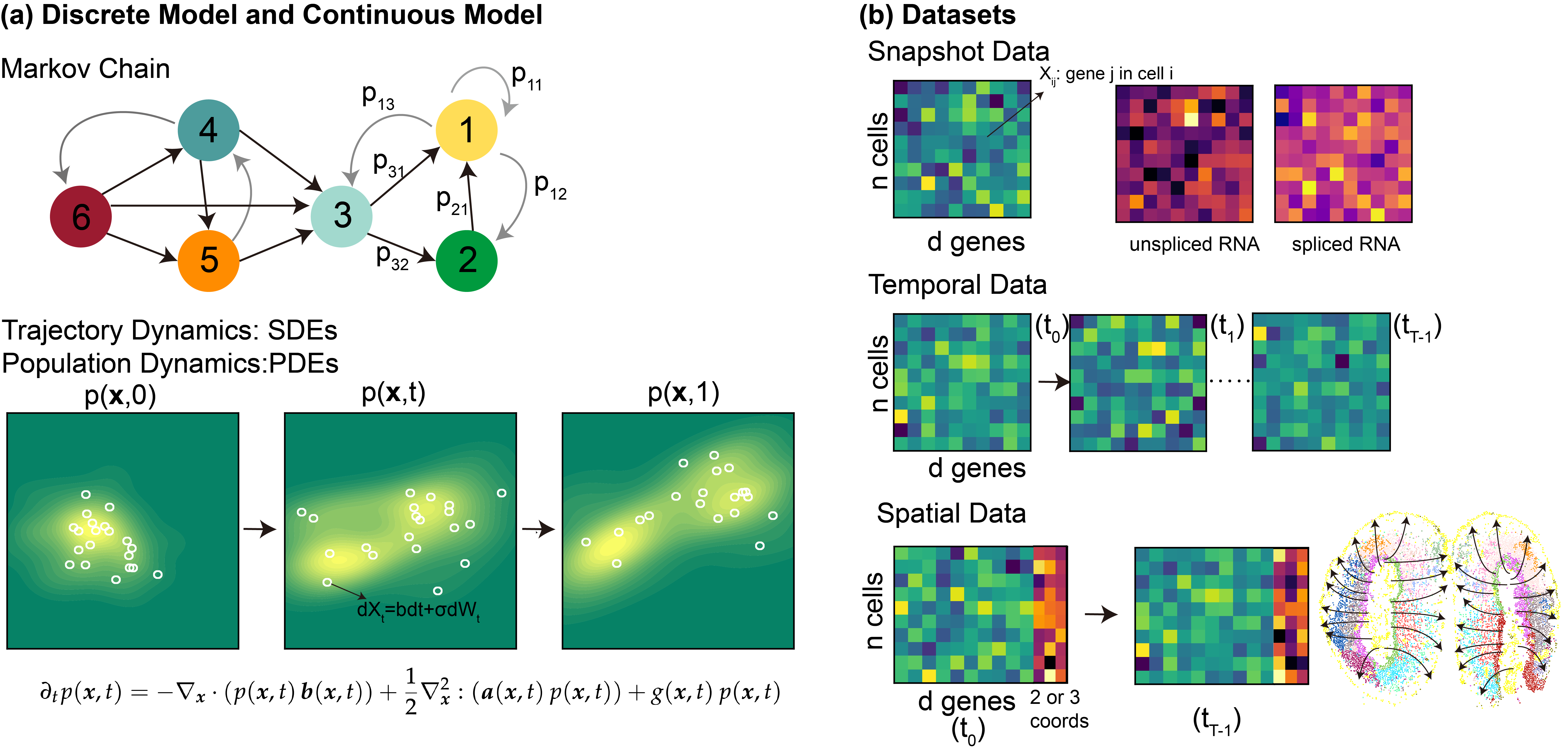}
\end{adjustwidth}
        \caption{\textbf{{Overview} 
 of the data and models.} ({\textbf{a}}) {Discrete and Continuous Model}: The discrete model constructs Markov chains between cells with dynamics encoded in a transition matrix, while the continuous model describes single-cell motion via stochastic differential equations (SDEs) and cell population dynamics through a corresponding partial differential equation (PDE). ({\textbf{b}}) {Datasets}: Snapshot data is an $n \times g$ matrix {$\boldsymbol{X}$} 
 ($n$: cell count, $g$: gene count); temporal data provides gene expression matrices $\boldsymbol{X}_i$ at time points $i \in \{0,\ldots,T-1\}$; spatial data additionally records coordinates for each cell in $\boldsymbol{X}_i$.}
        \label{fig:0}

\end{figure}

In this section, we provide preliminary background on the structure of the scRNA-seq data as well as the mathematical models to describe dynamic cellular processes. {An overview of the data and models is provided in Figure~\ref{fig:0}.}

\vspace{-2pt}

\unskip

\subsection{Spatiotemporal scRNA-Seq~Data}
Single-cell RNA sequencing (scRNA-seq) has emerged as a prevalent tool for dissecting cellular heterogeneity by providing high-resolution snapshots of gene expression profiles at the individual cell level. Traditionally, scRNA-seq experiments capture only single-time-point data, that is, a~static “snapshot” of the cellular landscape.  Recent advances in technologies have enhanced the spatiotemporal resolutions of the datasets, enabling finer resolutions to investigate the underlying dynamic biological processes such as development, differentiation, and~disease progression. 
Below we describe the various types of scRNA-seq datasets as inputs to infer spatiotemporal dynamics through the dynamical systems~models.
\subsubsection{Snapshot scRNA-Seq~Data}
In snapshot RNA sequencing (RNA-seq) data, gene expression is measured across multiple cells at a single time point. The~gene expression matrix is represented as \(\boldsymbol{X} \in \mathbb{R}^{n \times d} \), where \(\boldsymbol{X} \) denotes the count matrix of gene expression, \(n \) is the number of cells or spots, and~\(d \) is the number of genes measured. Each entry \(X_{ij} \) in \(\boldsymbol{X} \) represents the expression level of gene \(j \) in cell \(i \), typically measured as the number of mRNA molecules (transcripts) for that gene in the corresponding cell.  Additionally, the~total RNA-seq data can further be separated into counts for spliced and unspliced transcripts, useful in certain analyses such as the RNA velocity model described below. The~spliced and unspliced counts are represented as \(\boldsymbol{U} \in \mathbb{R}^{n \times d} \) and \(\boldsymbol{S} \in \mathbb{R}^{n \times d} \), respectively, denoting the matrices of unspliced and spliced counts for the \(n \) cells or spots. Over~time, unspliced RNA (\(\boldsymbol{u} \)) can undergo splicing process to become spliced RNA (\(\boldsymbol{s} \)).

\subsubsection{Temporally and Spatially Resolved~scRNA-Seq}

Recently, a~growing number of temporally resolved scRNA-seq datasets have been generated, where single-cell measurements are performed at multiple time points during a dynamic process. Such datasets could offer deeper insights into how cell populations evolve over time~\cite{nature_review_gene_temporal, schiebinger2021reconstructing, bunne2024optimal}. 

For temporally resolved scRNA-seq dataset, at~each fixed time point \(i \in \{0, \dots, T-1\} \), the~gene expression matrix is represented as \(\boldsymbol{X}_i \in \mathbb{R}^{n_i \times d} \), where \(\boldsymbol{X}_i \) denotes the matrix of gene expression data, \(n_i \) is the number of cells at time \(i \), and~\(d \) is the number of genes. Notably, the~gene expression data across time points are unpaired and can be assumed to be sampled from a distribution from a certain time~point.

The development of spatial transcriptomics (ST) technology allows gene expression to be captured alongside spatial coordinates~~\cite{staahl2016visualization, rodriques2019slide, stickels2021highly, chen2022spatiotemporal, oliveira2024characterization, moffitt2018molecular, eng2019transcriptome, wang2018three}. ST methods are broadly divided into image-based and sequencing-based approaches. Image-based techniques~~\cite{moffitt2018molecular, eng2019transcriptome, wang2018three} detect hundreds to thousands of genes with cellular or sub-cellular resolution, while sequencing-based methods~~\cite{rodriques2019slide, stickels2021highly, chen2022spatiotemporal, oliveira2024characterization} allow whole-transcriptome analysis but are usually limited to spot-level resolution. Advances like Stereo-seq~\cite{chen2022spatiotemporal} and 10{x}  
 Visium HD~\cite{oliveira2024characterization} have significantly improved spatial resolution to single-cell or even subcellular~precision. 

Similarly to temporally resolved scRNA-seq data, ST time series data could be represented as $(\boldsymbol{Z}_{(0:K)},\boldsymbol{X}_{(0:K)})$ at $t_0, t_1 \dots t_K$ totaling $K$ time points, and~the number of cells in each observation is $n_0, n_1 \dots n_K$. In~addition to the gene expression matrices $\boldsymbol{X}_{i}\in \mathbb{R}^{n_{i}\times d}$, the~associated spatial coordinate matrices $\boldsymbol{Z}_{i}\in \mathbb{R}^{n_{i}\times 2}$ or $\mathbb{R}^{n_{i}\times 3}$ represent the spatial coordinates (2D or 3D) of each sequenced cell (or spot), respectively. 

\subsection{Models for Cell-State~Transitions}
In computational systems biology, several modeling strategies have been formulated to quantify the cell-state transition dynamics. In~general, they could be categorized into two types: discrete models, which are usually defined on observed samples and evolve in discrete time steps, as~well as continuous models, which are extrapolated into the continuous cell state space and described by differential equation~models. 

\subsubsection{Discrete Dynamics: Markov Chain~Model}
\textbf{Random walk} or \textbf{Markov chain} models are simple yet powerful tools for studying stochastic dynamic processes, particularly in the context of complex systems such as gene expression dynamics and cell trajectories. In~these models, a~system evolves over time as a series of transitions between discrete states, where each state corresponds to a possible configuration or position in the system (such as a specific gene expression profile or a cell's position in a developmental trajectory). The~transitions between states are governed by transition probabilities, which can be represented in the form of a transition matrix \(\boldsymbol{P} \). The~transition matrix is {defined as} 
\(P_{ij} = \frac{W_{ij}}{W_i},
\)
where \(P_{ij} \) represents the probability of transitioning from cell state \(i \) to state \(j \), and~\(W_{ij} \) is the weight (or similarity) between cells \(i \) and \(j \) in a weighted graph, and~$W_i=\sum_k W_{ik}$ is the degree of cell $i$. The~weights \(W_{ij} \) typically reflect some measure of similarity or distance between the corresponding gene expression profiles of the cells, or~induced from other quantities such as RNA velocity or optimal transport~plan. 

The \textbf{stationary distribution} \(\boldsymbol{\pi} = \left(\pi_1, \pi_2, \dots, \pi_n \right) \) of the Markov chain is a probability distribution over the states that remains unchanged under the dynamics of the chain. In~other words, the~distribution is invariant under the transition probabilities, and~we have
\(\pi_i = \sum_{j} P_{ij} \pi_j.
\)
If the cellular state graph is undirected (for example, induced by gene expression similarity), i.e.,~ $\boldsymbol{W}$ is a symmetric matrix and we have the expression$\pi_i=\frac{W_i}{\sum_i W_i}$, then the stationary Markov chain is in detailed balance such that  $P_{ij}\pi_i=P_{ji}\pi_j $.

A more realistic assumption in biology is that the state-transition graph can be directional (for example, induced by RNA velocity or optimal transport discussed below), with~the cells ultimately reaching terminal states such as fully differentiated or mature cell types. In~this setup, \textbf{recurrent states} represent the final, stable cell types or fates that the system eventually reaches. Once a cell enters one of these recurrent states, it remains there, similar to how a fully differentiated cell does not revert back to an undifferentiated or less specialized state. On~the other hand, \textbf{transient states} correspond to intermediate stages of cellular development, such as precursor or progenitor cells that are still undergoing differentiation or division. These cells are in transition, with~the potential to eventually reach one of the recurrent, stable cell types. The~transition matrix governing this system can be partitioned into blocks that reflect these different types of cell states. Specifically, the~matrix \(\boldsymbol{P} \) can be written as the {canonical form} 
\begin{equation}\label{eq:canonical}
\boldsymbol{P}=
\begin{bmatrix}
    {\boldsymbol{\tilde{P}}} & \boldsymbol{0} \\ 
    \boldsymbol{S} & \boldsymbol{Q}
\end{bmatrix}.    
\end{equation}
{Here,} 
 \(\tilde{\boldsymbol{P}} \) corresponds to the transitions between recurrent (terminal) states, where once a cell reaches these states, it remains there in absorbing states. \(\boldsymbol{Q} \) represents transitions between transient states (cells that are still in intermediate stages of differentiation or cell cycle). \(\boldsymbol{S} \) denotes the transitions from transient to recurrent states, representing cells' eventual differentiation or maturation into their terminal stable types. Since recurrent states are absorbing, the~upper right block of the matrix is zero, indicating no transitions from recurrent to transient~states.

\subsubsection{Continuous Dynamics: From Trajectories to Population~Dynamics}

In modeling cellular dynamics, we are interested in both the trajectories of individual cells and the distribution of cell states across a population. To~capture the behavior of cells in response to both deterministic and stochastic influences, we can approach the problem from two perspectives: (1) \textbf{cellular trajectories}, which describe the path of individual cells over time, and~(2) \textbf{population distribution}, which describes how the overall distribution of cell states evolves. The~first perspective, trajectory-based models, provides insight into the detailed behavior of a single cell, often described through ordinary or stochastic differential equations (ODEs or SDEs). The~second perspective, population-level models, focuses on the evolution of the density of cells across different states, typically captured by partial differential equations (PDEs). Together, these models offer a comprehensive understanding of how individual cell behaviors aggregate to produce population-level~dynamics.

\textbf{Trajectory Dynamics: Stochastic Differential Equations (SDEs)}
To model the evolution of individual cell trajectories, we consider that cellular dynamics can be governed by a stochastic differential equation (SDE). This accounts for both deterministic factors, such as gene expression regulation, and~stochastic factors, like noise from cellular environments or molecular fluctuations. The~SDE for the state \(\boldsymbol{x}_t \) of a single cell at time \(t \) is given by\vspace{-6pt}
\begin{equation}\label{eq:sde}
\mathrm{d} \boldsymbol{x}_t = \boldsymbol{b}(\boldsymbol{x}_t, t) \, \mathrm{d} t + \boldsymbol{\sigma}(\boldsymbol{x}_t, t) \, \mathrm{d} \boldsymbol{w}_t, 
\end{equation}
where \(\boldsymbol{x}_t \in \mathbb{R}^d \) represents the state of the cell (e.g., gene expression profile) at time \(t \), and~\(\boldsymbol{w}_t \in \mathbb{R}^d \) is the standard \(d \)-dimensional Brownian motion. The~term \(\boldsymbol{b}(\boldsymbol{x}, t) \) represents the drift vector, which defines the deterministic flow of the system, while \(\boldsymbol{\sigma}(\boldsymbol{x}, t) \in \mathbb{R}^{d\times d}\)  represents the diffusion coefficient matrix, which describes the random fluctuations in the~system. 

Specifically, when the diffusion coefficient \(\boldsymbol{\sigma}(\boldsymbol{x}, t) \) is zero, the~system reduces to an ordinary differential equation (ODE), which describes the deterministic evolution of the cell state without random fluctuations. In~this case, the~evolution of the cell is entirely governed by the drift term \(\boldsymbol{b}(\boldsymbol{x}, t) \), and~the system follows a deterministic trajectory. A~useful concept in understanding the long-term behavior of the system is that of an {\textit{attractor}.} 
~In~the context of the ODE, an~attractor corresponds to a stable fixed point of the system, where the rate of change \(\boldsymbol{b}(\boldsymbol{x}, t) \) of the cell state \(\boldsymbol{x} \) becomes zero. In~cellular dynamics, such attractors can represent stable gene expression profiles, such as differentiated or quiescent states, where the cell remains in a stable state over~time. 

Furthermore, since \(\boldsymbol{b}(\boldsymbol{x}, t) \) is time-dependent, the~system may exhibit \textit{bifurcations},
 where the qualitative characteristics of attractors could change with respect to time $t$. In~cellular contexts, bifurcations are important for understanding processes like cell fate decisions, where a small change in the environment or internal signaling can push the cell toward a new distinct state (e.g., differentiation into a different cell type). 

\textbf{{Population Dynamics: Partial Differential Equations (PDEs)}}
To capture the evolution of the entire population of cells, we consider the density of cells with respect to their state \(\boldsymbol{x} \), represented by the probability density function \(p(\boldsymbol{x}, t) \). The~population distribution evolves according to a partial differential equation (PDE), which incorporates both deterministic and stochastic dynamics at the population level. As~a simple model, the~evolution of \(p(\boldsymbol{x}, t) \) can be described by\vspace{-6pt}
\begin{equation}\label{eq:pde}
\partial_t p(\boldsymbol{x}, t) = -\nabla_{\boldsymbol{x}} \cdot \left(p(\boldsymbol{x}, t) \, \boldsymbol{b}(\boldsymbol{x}, t) \right) + \frac{1}{2} \nabla_{\boldsymbol{x}}^2: \left(\boldsymbol{a}(\boldsymbol{x}, t) \, p(\boldsymbol{x}, t) \right) + g(\boldsymbol{x}, t) \, p(\boldsymbol{x}, t),
\end{equation}
where \(\nabla_{\boldsymbol{x}}^2 :\left(\boldsymbol{a}(\boldsymbol{x}, t) \, p(\boldsymbol{x}, t) \right) = \sum_{ij} \partial_{ij} \left(a_{ij}(\boldsymbol{x}, t) p(\boldsymbol{x}, t) \right) \), and~\(\boldsymbol{a}(\boldsymbol{x}, t) = \boldsymbol{\sigma}(\boldsymbol{x}, t) \boldsymbol{\sigma}^T(\boldsymbol{x}, t) \) represents the diffusion matrix at the population~level. 

The terms on the right-hand side of the equation represent the key dynamics driving the population evolution. The~\textit{drift term} \(\nabla_{\boldsymbol{x}} \cdot \left(p(\boldsymbol{x}, t) \, \boldsymbol{b}(\boldsymbol{x}, t) \right) \) quantifies the deterministic flow of the population, describing how cells move through different states based on the drift vector \(\boldsymbol{b}(\boldsymbol{x}, t) \). The~\textit{diffusion term} \(\frac{1}{2} \nabla_{\boldsymbol{x}}^2: \left(\boldsymbol{a}(\boldsymbol{x}, t) \, p(\boldsymbol{x}, t) \right) \) models the spread of the population due to random fluctuations, where \(\boldsymbol{a}(\boldsymbol{x}, t) = \boldsymbol{\sigma}(\boldsymbol{x}, t) \boldsymbol{\sigma}^T(\boldsymbol{x}, t) \) represents the diffusion matrix, capturing the effects of stochasticity. Finally, the~\textit{ growth term} \(g(\boldsymbol{x}, t) \, p(\boldsymbol{x}, t) \) governs the birth and death rates of cells, modeling cell proliferation and mortality, thus controlling the population's size and dynamics over time. When \(g(\boldsymbol{x}, t) = 0 \), the~PDE reduces to the Fokker–Planck equation associated with the SDE in the Itô integral sense, describing the evolution of the probability density \(p(\boldsymbol{x}, t) \) for the stochastic process defined by the~SDE.

\section{Dynamic Modeling of Single-Cell~Transcriptomics}\label{sec:snaphots}
In this section, we describe the dynamical systems models for scRNA-seq datasets. We begin with methods for snapshot data, i.e.,~cells sequenced at a single time point, including pseudotime methods, discrete Markov chain methods, and~continuous RNA velocity methods and their extensions. Next, we review methods targeted for temporally resolved scRNA-seq data, majorly based on various formulations and extensions of optimal transport (OT) based~methods.

\subsection{ Snapshot Single-Cell~RNA-Seq}
A key challenge in using snapshot data to infer dynamic cellular trajectories lies in the inability to directly observe the temporal evolution of cells. The~destructive nature of the measurement process, where cells are disassociated after sequencing, means that we lack direct access to the temporal trajectory of individual~cells.  

\begin{figure}[H]

\begin{adjustwidth}{-\extralength}{0cm}
\centering 
        \includegraphics[width=0.95\linewidth]{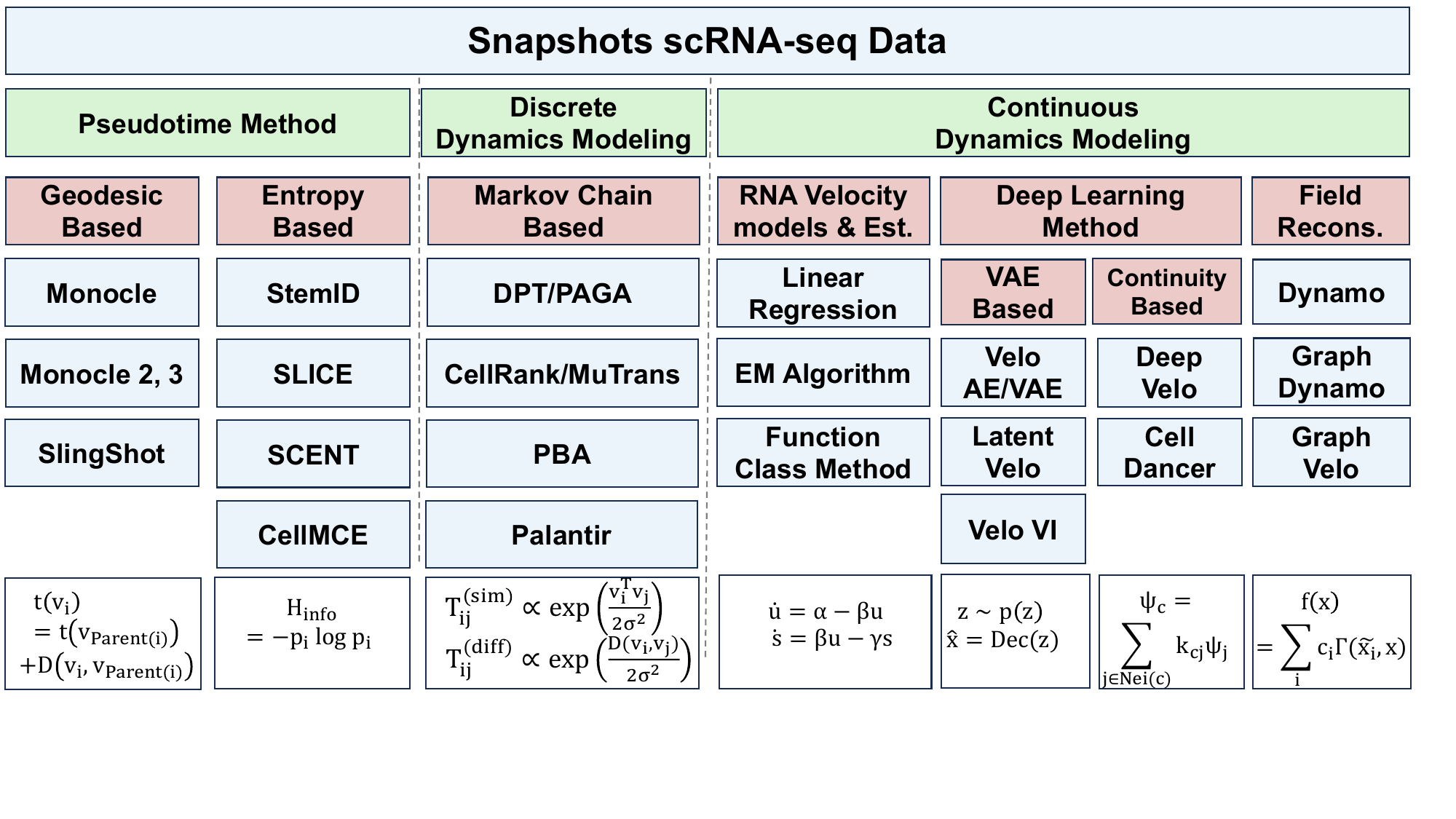}
\end{adjustwidth}
        \caption{\textbf{{Dynamic} 
 modeling of snapshot single-cell transcriptomics}.}
        \label{fig:1}

\end{figure}

When analyzing such a snapshot of ``cell state ensembles'', several approaches have been developed to uncover the underlying dynamical processes. One popular type of method, \textbf{pseudotime}, ranks individual cells temporally based on the structure of the data manifold or prior biological knowledge. Other techniques focus on modeling stochastic dynamics over the point clouds of observed cells, yielding \textbf{discrete random walk analyses}. Additionally, \textbf{continuous differential equation models} have been proposed to infer the data-generating process of snapshot scRNA-seq dataset, to~use these models for future dynamical predictions. In~the following sections, we will explore these methods in more detail. {An overview of these approaches is provided in Figure~\ref{fig:1}.}

\vspace{-3pt}
\unskip

{\subsubsection{Pseudotime~Methods} 
Given snapshot data from single-cell sequencing, where the data matrix $\textbf{X}\in\mathbb{R}^{n\times d}$, pseudotime assigns a positive real number for each cell to reflect its order during a dynamical process. Let \(\boldsymbol{x}_i \in \mathbb{R}^d \) represent the state vector (such as mRNA expression) for the \(i \)-th cell. The~pseudotime \(t_i \in \mathbb{R} \) is then a mapping from the state vector \(\boldsymbol{x}_i \) to a real number, i.e.,~\(\boldsymbol{x}_i \mapsto t_i \).

Pseudotime can be viewed from two perspectives. First, the~\textbf{geodesic} perspective considers that the cell's state is constrained by a limited number of biological pathways, limiting the evolution of the state to a low-dimensional manifold embedded in the high-dimensional gene expression space. Given an initial state \(\boldsymbol{x}_0 \), the~task of finding the mapping \(\boldsymbol{x}_i \mapsto t_i \) becomes equivalent to determining the length of the evolutionary path between \(\boldsymbol{x}_0 \) and \(\boldsymbol{x}_i \) along this manifold. Several tools, such as Monocle, Slingshot, DPT, and~PAGA, have been developed based on this idea. Second, the~\textbf{entropy} perspective recognizes that during natural biological development, as~cells differentiate, they become more specialized and lose their potential for further~differentiation. 

{\textbf{Geodesic-Based Pseudotime}} {Geodesic} 
-based pseudotime aims to reconstruct these trajectories by leveraging graph-based methods and principal curve algorithms~\cite{deconinck2021recent}. Two widely used approaches include \textbf{Monocle}~\cite{qiu2017monocle,cao2019singlemonocle3} and \textbf{Slingshot}~\cite{street2018slingshot}, along with numerous other methods. Monocle focuses on ordering cells using a minimum spanning tree and a refined PQ tree approach, and~Slingshot can handle multiple lineages and smooth pseudotime across branching events. Both methods offer insights into cellular dynamics, helping to uncover the paths cells take through different~states.

{Monocle} estimates pseudotime through two main steps: (1) ordering cells and \linebreak (2) assigning pseudotime values. First, each cell's state is reduced to a \(d \)-dimensional vector \(\boldsymbol{x}_i \in \mathbb{R}^d \) using Independent Component Analysis (ICA). A~complete graph is created where vertices represent cells, and~edges are weighted by Euclidean distance. Cell ordering relies on the minimum spanning tree (MST) of this graph, which is refined using a PQ tree to mitigate noise from sequencing. The~tree is constructed by first identifying the longest path (diameter path), classifying vertices as decisive or indecisive, and~recursively building the tree by ordering decisive vertices and handling indecisive vertices through new P nodes. Once ordered, pseudotime is calculated~as\vspace{-6pt}
\[
t(\boldsymbol{x}_i) = t(\boldsymbol{x}_{\text{Parent}(i)}) + \|\boldsymbol{x}_i - \boldsymbol{x}_{\text{Parent}(i)}\|,
\]
where \(\text{Parent}(i) \) is the parent node of cell \(i \), and~the root node (selected based on prior knowledge) is initialized with pseudotime~0.

{Slingshot} estimates pseudotime across multiple lineages. It begins by clustering cells into \(K \) clusters and identifying lineages using the MST.  After~constructing the MST, a~new lineage is formed at each branching point. Pseudotime is assigned using the principal curve algorithm, which involves projecting cells onto the curve, computing arc lengths, and~smoothing iteratively.
To handle inconsistent pseudotime across multiple lineages, Slingshot modifies the standard principal curve approach. It initializes the curve for each lineage through the centroids of its clusters, assigns weights to cells in multiple lineages based on projection distances, and~constructs an average curve for smooth transitions at shared cell regions. The~average curve is defined as\vspace{-6pt}
\[
\boldsymbol{c}_{\text{avg}}(t) = \frac{1}{M} \sum_{m=1}^M \boldsymbol{c}_m(t),
\]
where \(\boldsymbol{c}_m \) is the principal curve of the \(m \)-th lineage at the branching point. The~shrinkage process is defined as

\[
\boldsymbol{c}_m^{\text{new}}(t) = w_m(t) \boldsymbol{c}_{\text{avg}}(t) + (1 - w_m(t)) \boldsymbol{c}_m(t),
\]
where \(w_m(t) \) is the weight of the \(m \)-th lineage. These modifications allow Slingshot to produce consistent pseudotime values across multiple~lineages.

{\textbf{Entropy-Based Pseudotime}}
One major challenge of the geodesic-based pseudotime is the appropriate determination of root cells, which often relies on prior biological knowledge. From~a physical understanding, a~cell's pseudotime reflects the directionality of the underlying dynamical process, which the concept of entropy could quantify. Heuristically, higher entropy values typically indicate a more undifferentiated or pluripotent state where genes are more randomly expressed and the association between genes could be more prevalent. In~comparison, lower entropy values suggest differentiated states, where the gene expression profile could be more concentrated on only a small number of pathways, and~the gene interaction network could be more modular~\cite{gandrillon2021entropy}. A~proper entropy score based on such intuitions can thus be leveraged to estimate a cell's relative position along developmental trajectories. Several methods have been developed from this perspective~\cite{grun2016novo,guo2017slice,teschendorff2017single,shi2020quantifying,jin2018scepath,liu2020single}.

As a simple implementation, the~entropy for one cell \(i \) is defined as~\cite{grun2016novo}\vspace{-6pt} $$
H_{i} = -\sum_{j=1}^{d} p_{ij} \log p_{ij},
$$
where \(p_{ij} = \frac{X_{ij}}{N_{i}} \), and \(X_{ij} \) represents the transcript count of gene \(j \) in cell \(i \), and~\(N_{j} \) is the total transcript count for cell \(i \).

One can extend this concept to a Markov chain model~\cite{teschendorff2017single} to consider the interaction between the genes. Assume there is a predefined graph representing the gene–gene interaction (e.g., protein–protein interaction (PPI) network) from the existing database. The~transition probability between genes $i$ and $j$ in cell \(c \) is
$$
p_{ij}^{(c)} = \frac{x_{j}^{(c)}}{\sum_{k \in N(i)} x_{k}^{(c)}} = \frac{x_{j}^{(c)}}{(\boldsymbol{A}\boldsymbol{x}^{(c)})_{i}},
$$
where \(x_{i}^{(c)} \) is the expression level of gene \(i \) in cell \(c \), \(N(i) \) are the neighbors of gene \(i \) in the graph, and~\(\boldsymbol{A} \) is the adjacency matrix of the graph. The~corresponding stationary distribution is
$$
\pi^{(c)}_{i} = \frac{x^{(c)}_{i}(\boldsymbol{A}\boldsymbol{x}^{(c)})_{i}}{\boldsymbol{x}^{(c)^T}\boldsymbol{A}\boldsymbol{x}^{(c)}},
$$
and the Markov chain entropy (MCE) is defined as 
$$
\text{MCE} = - \sum_{(i,j) \in \widetilde{E}} \pi_{i} p_{ij} \log(\pi_{i} p_{ij}),
$$
where \(\widetilde{E}\) includes all edges on the graph. The~entropy of cell \(c \) is given by \(\text{MCE}^{(c)} \), computed using \(\pi_{i}^{(c)} \) and \(p_{ij}^{(c)} \).
To determine the weights, {ref.}~\cite{shi2020quantifying} proposes to optimize interaction weights based on cell expression \(\boldsymbol{\pi}^{(c)} = \frac{\boldsymbol{x}^{(c)}}{\|\boldsymbol{x}^{(c)}\|_{\mathcal{L}_{1}}} \). For~each cell, its MCE is maximized by~solving
\[
 \max_{p_{ij}^{(c)} \ge 0} - \sum_{(i,j) \in \bar E} \pi_{i}^{(c)} p^{(c)}_{ij} \log (\pi^{(c)}_{i} p^{(c)}_{ij}),
\]
subject to
\(\sum_{j \in \mathcal{N}(i)} p^{(c)}_{ij}= 1 \quad \text{and} \quad \sum_{i \in \mathcal{N}(j)} \pi^{(c)}_{i}p^{(c)}_{ij} = \pi_{j}^{(c)}.
\)
{\subsubsection{Discrete Dynamics~Modeling}}
{\textbf{Diffusion Pseudotime}} 
Previous pseudotime methods, such as those based on geodesic paths or simple assumptions of pseudotime, typically lacked an underlying dynamical model to explain cell state transitions. These methods often relied on the assumption of continuous trajectories without explicitly modeling the stochastic processes driving those transitions. In~contrast, methods like \textbf{DPT}~\cite{haghverdi2016diffusion} and \textbf{PAGA}~\cite{wolf2019paga} introduce stochastic dynamics through Markov chains. While still estimating pseudotime, these methods incorporate random walk-defined observed samples of single cells, allowing for a more quantitative treatment of how cell states evolve over time, with~transitions captured probabilistically. As~a result, they provide a more mechanistic approach to pseudotime estimation, making them a natural progression from traditional geodesic-based~methods.

Motivated by the DiffusionMap~\cite{coifman2006diffusion} algorithm for dimensionality reduction, {DPT} constructs a Markov chain between cells, defines a distance metric, and~uses this distance as pseudotime. The~transition probability of cell \(i \) moving to cell \(j \) is computed using a simple Gaussian kernel, defined as\vspace{-6pt}
\[
T_{ij} = \frac{1}{Z} W_{ij} = \frac{1}{Z} \left(\frac{2 \sigma_{i} \sigma_{j}}{\sigma_{i}^{2} + \sigma_{j}^{2}} \right) \exp \left(- \frac{\|\boldsymbol{x}_{i} - \boldsymbol{x}_{j}\|^{2}}{2 (\sigma_{i}^{2} + \sigma_{j}^{2})} \right),
\]
where \(Z_{i} = \sum_{j \in \mathcal{N}(i)} W_{ij} \) is the normalization factor, and~the hyperparameters \(\sigma_i, \sigma_j \) are the Gaussian kernel widths for cell \(i \) and cell \(j \). DPT assumes that the distance in the eigenspace of the transition matrix \(T \) is related to the pseudotime ordering of cells. After~removing steady-state eigenspace, the~system's dynamics are captured by the transition matrix \linebreak \(\bar{\boldsymbol{T}} = \boldsymbol{T} - \boldsymbol{\psi}_{1} \boldsymbol{\psi}_{1}^{T} \), where  \(\boldsymbol{\psi}_{1} \) represent the eigenvector corresponding to the largest eigenvalue of the transition matrix. The~dynamics are analyzed by summing all \(t \)-step transition matrices to compute the cumulative probability of state transitions across multiple walk~lengths\vspace{-6pt}
\[
\boldsymbol{M} = \sum_{t=1}^{\infty} \bar{\boldsymbol{T}}^{t} = (1 - \bar{\boldsymbol{T}})^{-1} - \boldsymbol{I}.
\]
Using this matrix \(\boldsymbol{M} \), a~new distance metric is defined as
\[
\text{dpt}^{2}(i,j) = \|\boldsymbol{M}_{i,\cdot} - \boldsymbol{M}_{j,\cdot}\| = \sum_{k=2}^{n} \left(\frac{\lambda_{k}}{1 - \lambda_{k}}\right)^{2} ({\psi}_{k}(i) - {\psi}_{k}(j))^2,
\]
where \({\psi}_{k}(i) \) represents the \(i \)-th component of the eigenvector corresponding to the \(k \)-th largest eigenvalue of the transition matrix. This metric simultaneously captures both short-range and long-range cell state transitions, making it useful for understanding the trajectory of cell states over~time.

{PAGA} generalizes the DPT distance metric to disconnected graphs to deal with the existence of multiple distinct lineages in the dataset. In~PAGA, graph construction begins by reducing the dimensionality of the gene expression data using PCA, followed by the construction of a KNN graph where the nodes represent cells. The~graph is then partitioned into cell clusters using the Louvain algorithm, reminiscent of the \textbf{attractors} concept in the random walk. Two groups are considered connected if the actual number of edges \(\epsilon_{ij} \) between them significantly exceeds the expected number of edges. The~DPT distance metric is then extended to the disconnected graph. In~practice, one treats cells that belong to separate clusters as being at an infinite distance from each other. For~cells within the same connected region, one calculates distances between them similar to the calculation in DPT. This modification allows PAGA to estimate pseudotime and infer trajectories even in the presence of disconnected or sparse data~regions.

{\textbf{Random Walk with Directionality}}
Building on the inferred DPT, Palantir~\cite{setty2019characterization} introduces directionality into the cellular random walk, which is further developed in~\cite{stassen2021generalized,pandey2022inference}. One simple approach is to prune the weight matrix as follows:\vspace{-6pt}
\[
\bar{W}_{ij} =
\begin{cases}
W_{ij} & \text{if } t_i \leq t_j \text{ or } 0 < t_i - t_j < \sigma_i \\
0 & \text{if } t_i - t_j > \sigma_i
\end{cases}
\]
For the directional Markov chain induced by the weight matrix \(\bar{\boldsymbol{W}} \), terminal states can be determined from its stationary distribution. Using absorption Markov chain theory, a~cell fate matrix \(\boldsymbol{F} \) can be derived from the canonical form of transition probability matrix in Equation \eqref{eq:canonical}, where \(\boldsymbol{F} = (\boldsymbol{I} - \boldsymbol{Q})^{-1} \boldsymbol{S} \). Here, the~element \(\boldsymbol{F}_{ij} \) represents the probability that a random walk starting from transient cell \(i \) will eventually be absorbed by terminal cell \(j \). Specifically, the~fate vector \(\boldsymbol{f}_i \) of a transient cell \(i \) corresponds to the \(i \)-th row of \(\boldsymbol{F} \), capturing the probability of differentiation into various states. To~quantify the differentiation potential of a cell, one can then determine the entropy of the fate vector, or~the Kullback–Leibler (KL) divergence between the fate vector \(\boldsymbol{f}_i \) and the average fate vector \(\boldsymbol{\bar{f}} \). Author: Please check that the intended meaning has been~retained.

Another method to define the directional random walk on cells is Population Balance Analysis (PBA)~\cite{weinreb2018fundamental}.   Let $\boldsymbol{G}$ be the $k$-nearest neighbor graph of $\{\boldsymbol{X}\}$ and $\boldsymbol{L}$ its graph Laplacian. A~potential function is defined by  $\boldsymbol{V} = \dfrac{1}{2} \boldsymbol{L}_{N}^{\dagger} \boldsymbol{R}$, where $\boldsymbol{L}_{N}^\dagger$ denotes the pseudo-inverse of $L_{N}$ and $\boldsymbol{R}$ is the estimated cell population production rate vector at each node, using the gene expression of predefined lists of proliferation relevant genes. The~transition probabilities of this Markov chain are then directed by the potential function\vspace{-6pt}
$$
P_{ij}= 
\begin{cases}
\exp\left(\dfrac{\boldsymbol{V}_{i} - \boldsymbol{V}_{j}}{D}\right) \quad  &\text{if}\ (i,j)\  \text{is in } \boldsymbol{G}_{N} \\
0 &\text{otherwise}
\end{cases}
$$
After the random walk is constructed, PBA utilizes the conditional mean first passage time to quantify the difference of pseudotime 
for any pair of transient cells $i$ and $j$.

{\textbf{Dissecting Dynamical Structure}}
While previous methods such as DPT, Palantir, and~PBA construct the random walk dynamics on snapshots of individual cells, methods like \textbf{MuTrans}~\cite{zhou2021dissecting} and \textbf{CellRank}~\cite{lange2022cellrank, weiler2024cellrank} take a deeper look into the dynamical structure of the system itself, especially focusing on \textbf{metastability} and \textbf{attractor} structures, therefore robustly dissecting the system's latent dynamics to identify long-term patterns of the cell-state~transition. 

{MuTrans}~\cite{zhou2021dissecting} adopts a multi-scale reduction technique for a diffusion-based, unidirectional cellular random walk, to~infer stable and transient cells from snapshot scRNA-seq. Central to MuTrans is a membership matrix \(\chi_{i,k} \) representing the soft clustering probability that cell \(i \) belongs to attractor $S_k,k=1,\ldots,K$, which could be interpreted as a cell cluster. A~transient cell might have multiple positive components in its attractor membership, while the distribution of stable cells tends to be concentrated in one specific attractor.  Meanwhile, MuTrans also reduces the dynamics on the attractor level, using \(\boldsymbol{P}^{(\text{coar})}\in\mathbb{R}^{K\times K} \) to represent the coarsened transition matrix, and~\(\boldsymbol{\pi}^{(\text{coar})} \) to represent the stationary distribution of the coarse-grained Markov chain. The~original cell-to-cell dynamics can then be reconstructed from the coarsened dynamics, and~the transition probability between cells \(i \) and \(j \) is given~by\vspace{-6pt}
\[
\hat{P}_{i,j} = \sum_{m,n=1} ^{K}\chi_{i,m} P^{(\text{coar})}_{m,n} \chi_{j,n} \frac{{\pi}_{j}}{\pi^{(\text{coar})}_n}.
\]
The goal is to minimize the discrepancy between the reconstructed cell–cell dynamics \(\hat{\boldsymbol{P}} \) and the actual dynamics \(\boldsymbol{P} \), which is achieved by minimizing \(\|\hat{\boldsymbol{P}} - \boldsymbol{P}\|^2 \). This can be done using an EM-like algorithm, alternately optimizing the elements of \(\boldsymbol{P}^{(\text{coar})} \) and \(\chi_{i,j} \). With~the inferred attractor membership matrix and coarse-grained transition probabilities among clusters, MuTrans then constructs a dynamical manifold inspired by the energy landscape concept~\cite{landscape_PZ} to visualize the transient and stable cells, and~uses transition path theory~\cite{vanden2010transition} based on ${P}^{(\text{coar})}$ to calculate the most probable transition paths among~attractors.

{CellRank} extends the analysis by introducing a coarse-graining strategy for directed cellular random walks, such as those induced by pseudotime~\cite{setty2019characterization} or RNA velocity as described in Section~\ref{sec:bridge}.  The~approach begins with the clustering of cells into macro-states (i.e., attractors) using the GPCCA (Generalized Perron Cluster Cluster Analysis) algorithm~\cite{reuter2019generalized}, which is based on the Shur decomposition of the directed transition matrix \(\boldsymbol{P} \). The~membership matrix $\boldsymbol{\chi}$ and the coarse-grained transition probability matrix between attractors $\boldsymbol{P}^{(\text{coar})}$ can be computed based on the decomposition. Once the attractors are identified, the~terminal states can be determined in which the diagonal elements of the coarse transition matrix \(\boldsymbol{P}^{(\text{coar})} \) exceed a certain threshold.  Cells in terminal states can then be treated as absorption sets of the random walk, and~the cell fate vector could be computed similarly to Palantir.
\vspace{1em}
{\subsubsection{Continuous Dynamics~Modeling}}
{\textbf{RNA Velocity Model and Parameter Estimation}}
Based on the unspliced RNA and spliced RNA counts $u_g$ and $s_g$ for each gene, an~underlying ODE model could be naturally derived based on mass-action law such that
\begin{equation}
    \begin{aligned}
        \dfrac{\mathrm{d} u_g}{\mathrm{d} t} &=  \alpha_g(t)  -  \beta_g(t) u_g(t),  \\ 
        \dfrac{\mathrm{d} s_g}{\mathrm{d} t} &=  \beta_g(t) u_g(t) - \gamma_g(t) s_g(t),
    \end{aligned}
    \label{velocityeq}
\end{equation}
where $\alpha_g(t)$, $\beta_g(t)$, and~$\gamma_g(t)$ represent the rates of mRNA transcription, splicing, and~degradation, respectively. Here, $v_g = \dfrac{\mathrm{d} s_g}{\mathrm{d} t}$ is defined as the \textbf{RNA velocity} of gene $g$~\cite{la2018rna}. By~concatenating the RNA velocities of all genes in a cell, a~vector $\boldsymbol{v} = (v_1, v_2, \cdots, v_n)$ is formed, which contains information about how the amounts of spliced RNA in the cell are changing. This vector represents the potential direction of the cell state evolution and can be used for downstream tasks of cell fate~inference. 

From an algorithm perspective, the~central issue in the RNA velocity is to determine the parameters of the Equation \eqref{velocityeq} from static snapshot data, where the time $t$ of each cell is not explicitly known.  In~the following sections, we will summarize methods for solving $v_g$ and downstream tasks that utilize $v_g$.

\vspace{12pt}
\paragraph{{Steady-State} 
 Assumption: Parameter Estimation in Velocyto~\cite{la2018rna}}

In the original RNA velocity paper~\cite{la2018rna}, parameter estimation was performed using linear regression with steady-state assumption to avoid the reliance on latent time $t$. Firstly, it assumes that for all genes $g$, $\alpha_g(t)$, $\beta_g(t)$, and~$\gamma_g(t)$ are time-invariant. Secondly, it assumes that all genes share the same splicing rate $\beta$. Denote $\tilde{\alpha} = \dfrac{\alpha}{\beta}$ and $\tilde{\gamma} = \dfrac{\gamma}{\beta}$. To~compute RNA velocity, one only needs to estimate $\tilde \gamma_g$ with the steady-state assumption that $\frac{ds_g}{dt}=0$. Indeed, we have the linear relation  $\tilde \gamma_g = \dfrac{u_g(t)}{s_g(t)},$ suggesting that under the steady-state assumption, $\tilde \gamma_g$ can be estimated using linear regression. In~practice, since most cells do not satisfy this assumption, it is commonly assumed that cells in the upper-right or lower-left regions of a scatter plot with unspliced RNA on the x-axis and spliced RNA on the y-axis are in equilibrium. Therefore, the~common algorithm implementation is to limit the linear regression to the top or bottom $5\%$ of cells based on unspliced and spliced RNA~levels.

If the dynamic equations Equation~(\ref{velocityeq}) are expressed probabilistically, the~parameter estimation could be enhanced based on the linear regression formulation of steady-state stochastic model~\cite{bergen2020generalizing}. The~equation could be expressed as the regression problem\vspace{-6pt}
$$
    \begin{bmatrix}
    \langle u_g(t) \rangle \\ 
    \langle u_g(t) \rangle +  2 \langle u_g(t) s_g(t) \rangle
    \end{bmatrix} 
    = 
    \tilde \gamma_g \begin{bmatrix}
        \langle s_g(t) \rangle \\ 
        2\langle s_g^2(t) \rangle - \langle s_g(t) \rangle
    \end{bmatrix} + \boldsymbol{\epsilon},
$$
where $\langle x\rangle$ denotes the expectation of random variable $x$. The~regression equation incorporates both first-order and second-order moment information of $u_g(t)$ and $s_g(t)$ and can be solved using generalized least~squares.

\paragraph{{Dynamic} Inference: Parameter Estimation in scVelo~\cite{bergen2020generalizing} }

A major issue with steady-state analysis is that many transient cells would be discarded in the parameter estimation. The~scVelo approach~\cite{bergen2020generalizing} circumvents the issues through the estimation of kinetic parameters ($\alpha_g$, $\beta_g$, $\gamma_g$) via an expectation-maximization (EM) algorithm by modeling all the cells with the dynamical~process.

Guided by transcriptional regulation principles, \textbf{scVelo} models gene expression dynamics through two distinct transcriptional phases: (1) an \textit{induction phase} ($k=0$) characterized by promoter activation and transcriptional upregulation, and~(2) a subsequent \textit{repression phase} ($k=1$) marked by transcriptional suppression. This phase-specific regulation manifests through different unspliced RNA production rates ($\alpha^{(0)}_g \neq \alpha^{(1)}_g$). Let $t^{(k)}_g$ denote the transition time from phase $k-1$ to $k$ for gene $g$, with~initial conditions $u_g^\star = u_g(t^{(k)}_g)$ and $s_g^\star = s_g(t^{(k)}_g)$. The~analytical solution to  Equation~(\ref{velocityeq}) during phase \linebreak $k$ yields\vspace{-6pt}
$$
    \begin{aligned}
        u_g(t) &= u_g^\star e^{-\beta_g \tau} + \frac{\alpha_g^{(k)}}{\beta_g}(1 - e^{-\beta_g \tau}), \\ 
        s_g(t) &= s_g^\star e^{-\gamma_g \tau} + \frac{\alpha_g^{(k)}}{\gamma_g}(1 - e^{-\gamma_g \tau}) 
                + \frac{\alpha_g^{(k)} - \beta_g u_g^\star}{\gamma_g - \beta_g} (e^{-\gamma_g \tau} - e^{-\beta_g \tau}),  \tau &= t - t^{(k)}_g.
    \end{aligned}
$$

For each gene $g$, one can estimate the parameter set $\boldsymbol{\theta}_g = \{\alpha^{(k)}_g, \beta_g, \gamma_g, t^{(k)}_g\}$ by minimizing the discrepancy between modeled trajectories $\boldsymbol{\hat{x}}_g(t) = (u_g(t), s_g(t))$ and observed single-cell measurements $\boldsymbol{x}_{g,c} = (u_{g,c}, s_{g,c})$ across cells $c$. Assuming Gaussian residuals $\boldsymbol{e}_c = \|\boldsymbol{x}_{g,c} - \boldsymbol{\hat{x}}_g(t_c)\|$ with variance $\sigma^2$, the~log-likelihood function becomes\vspace{-6pt}
\begin{equation}
    \max_{\boldsymbol{\theta_g}, t_c}\mathcal{L}(\boldsymbol{\theta}_g,,t_c) = -\frac{1}{2\sigma^2}\sum_{c} \|\boldsymbol{x}_{g,c} - \boldsymbol{\hat{x}}_g(t_c)\|^2 + \text{constant}.
    \label{loglikelihood}
\end{equation}

The EM implementation proceeds as follows:

\begin{itemize}
    \item \textbf{Initialization:} Using steady-state estimation as the initial value for iteration. \vspace{-6pt}
    \begin{align*}
        \beta_g &= 1, \quad \gamma_g = \frac{\boldsymbol{u}_g^\top \boldsymbol{s}_g}{\|\boldsymbol{s}_g\|^2}, \\
        k_{g,c} &= \mathbb{I}(u_{g,c} - \tilde{\gamma} s_{g,c} \geq 0), \quad \alpha_g^{(1)} = \max_c s_{g,c}, \quad \alpha^{(0)}_g = 0.
    \end{align*}
    
    \item \textbf{E-step:} Assigning hidden latent time $t_{c}$ for each cell by projecting observations onto the current estimated trajectory  $\boldsymbol{\hat{x}}_g(t|\boldsymbol{\theta}_g)$.
    
    \item \textbf{M-step:} Updating $\boldsymbol{\theta}_g$ via maximum likelihood estimation given current latent time assignments.
\end{itemize}

\paragraph{{Function} Class-Based Estimation}

While traditional RNA velocity methods focus on estimating parameters $\alpha_g, \beta_g, \gamma_g$ in dynamic Equation \eqref{velocityeq}, alternative approaches such as UniTVelo~\cite{gao2022unitvelo} and TF Velo~\cite{li2024tfvelo} take a different path by directly parameterizing the dynamics of spliced RNA.  UniTVelo~\cite{gao2022unitvelo} models transcriptional phases through radial basis functions
$$
    \begin{aligned}
        s_g(t_{g,c}) &= h_g \exp\big(-a_g (t_{g,c} - \tau_g)^2\big) + o_g, \\ 
        u_g(t_{g,c}) &= \frac{1}{\beta_g}\left(\dot{s}_g(t_{g,c}) + \gamma_g s_g(t_{g,c})\right) + i_g,
    \end{aligned}
$$where the velocity derives directly from function differentiation
$
    \dot{s}_g(t_{g,c}) = -2a_g(t_{g,c}-\tau_g)s_g(t_{g,c}).
$
The full parameter set $(h_g, a_g, \tau_g, o_g, \gamma_g, \beta_g, i_g, t_{g,c})$ is estimated via maximum likelihood framework as Equation \eqref{loglikelihood}, comparing model predictions $\boldsymbol{\hat{x}}_c = (u_g(t_{g,c}), s_g(t_{g,c}))$ against observations $\boldsymbol{x}_c = (u_{g,c}, s_{g,c})$ under Gaussian~residuals.

TF Velo~\cite{li2024tfvelo}  introduces transcription factor coupling through linear dynamics
$
    \dot{s}_g(t) = \boldsymbol{w}_g^\top \boldsymbol{f}_g(t) - \gamma_g s_g(t),
$
combined with sinusoidal splicing
$
    s_g(t) = A_g \sin(\omega_g t + \theta_g) + b_g.
$
This functional specification enables the analytical resolution of TF interactions\vspace{-6pt}
$$
    \boldsymbol{w}_g^\top \boldsymbol{f}_g(t) = A_g\sqrt{4\pi^2+\gamma_g^2}\sin(2\pi t + \theta_g + \phi_g) + b_g\gamma_g, \quad \phi_g = \arctan(2\pi/\gamma_g).
$$
Parameters are optimized by matching predicted trajectories $\boldsymbol{\hat{x}}_c = (\boldsymbol{w}_g^\top \boldsymbol{f}_g(t_{g,c}), s_g(t_{g,c}))$ to observed data $\boldsymbol{x}_c = (\boldsymbol{w}_g^\top \boldsymbol{f}_{g,c}, s_{g,c})$ using the same likelihood framework as  Equation~(\ref{loglikelihood}).

{\textbf{Deep Learning-Based RNA Velocity}} Recently, the~application of deep learning methods has expanded the possibilities for RNA velocity estimation~\cite{oller2021algorithmic, raimundo2021machine, gu2022bayesian}.  In~RNA velocity analysis, the~expressive power of neural networks is especially useful in inferring the latent state of cells, as~well as encouraging the consistency of the learned vector~field.

\paragraph{{Latent} State: VAE-Based Methods}

The Variational Autoencoder (VAE)~\cite{kingma2013auto} is an effective approach to model the distribution of data through latent variables to achieve high-dimensional data reconstruction. Its core idea is to introduce a latent variable $\boldsymbol{z}$ and express the distribution of data $\boldsymbol{x}$ as the following conditional distribution:
$
p(\boldsymbol{x}) =  \int p(\boldsymbol{x}|\boldsymbol{z}) p(\boldsymbol{z}) \mathrm{d}\boldsymbol{z},
$
where the generative process is
$
\boldsymbol{z} \sim  p(\boldsymbol{z}), \quad \boldsymbol{x} \sim  p(\boldsymbol{x}|\boldsymbol{z}).
$
The VAE employs a decoder $p_\theta(\boldsymbol{x}|\boldsymbol{z})$ and an encoder $q_{\phi}(\boldsymbol{z}|\boldsymbol{x})$ to map between the latent space and data space. The~training objective is the Evidence Lower Bound (ELBO):\vspace{-6pt}
$$
\mathcal{L}_{\mathrm{ELBO}} = \mathcal{L}_{\mathrm{rec}} + \mathcal{L}_{\mathrm{reg}} = \mathbb{E}_{q_{\phi}(\boldsymbol{z}|\boldsymbol{x})}[\log p_\theta(\boldsymbol{x}| \boldsymbol{z})] - \mathrm{D}_{\mathrm{KL}}[q_{\phi}(\boldsymbol{z}|\boldsymbol{x}) \| p(\boldsymbol{z})], \quad \boldsymbol{x} \sim p_{\mathrm{data}}(\boldsymbol{x}),
$$
where the first term is the \textbf{reconstruction loss}, enforcing similarity between decoded data and observations, and~the second term acts as a \textbf{regularizer}, aligning the latent distribution with the prior. Typically, the~decoder outputs the mean $\boldsymbol{\mu}_\theta(\boldsymbol{z})$ of a Gaussian distribution with fixed variance $\sigma^2$, leading to
$
\mathcal{L}_{\mathrm{rec}} \propto - \dfrac{1}{2 \sigma^{2}} \mathbb{E}_{q_{\phi}(\boldsymbol{z}|\boldsymbol{x})} \left[ \|\boldsymbol{x} - \boldsymbol{\mu}_\theta(\boldsymbol{z})\|^{2} \right].
$
Several methods are based on VAE to improve the RNA velocity model by taking advantage of the latent~space.

{VeloAE}~\cite{qiao2021representation} computes RNA velocity using latent space representations. Its encoder maps the spliced RNA matrix $\boldsymbol{S} \in \mathbb{R}^{n_{c} \times n_{g}}$ and unspliced RNA matrix $\boldsymbol{U} \in  \mathbb{R}^{n_{c}\times n_{g}}$ to latent representations $\tilde{\boldsymbol{S}} \in \mathbb{R}^{n_{c} \times d_{z}}$ and $\tilde{\boldsymbol{U}}\in \mathbb{R}^{n_{c} \times d_{z}}$, while the decoder reconstructs $\hat{\boldsymbol{S}}$ and $\hat{\boldsymbol{U}}$. Velo AE enforces the steady-state constraint on latent representations $\boldsymbol{\tilde{u}}_{i},\boldsymbol{\tilde{s}}_{i}$, resulting in a composite loss:\vspace{-12pt}
$$
\mathcal{L} = \mathcal{L}_{\mathrm{rec}} + \mathcal{L}_{\mathrm{reg}} =  \sum_{i=1}^{d_{z}} \mathrm{MSE}(\boldsymbol{\tilde{u}}_{i}- \gamma_{i}\boldsymbol{\tilde{s}}_{i}) + \left[\mathrm{MSE}(\boldsymbol{S}, \boldsymbol{\hat{S}}) + \mathrm{MSE}(\boldsymbol{U},\boldsymbol{\hat{U}})\right].
$$
The RNA velocity is derived as $\boldsymbol{\tilde{u_{i}}} - \gamma_{i}\boldsymbol{\tilde{s_{i}}}$ after~training.

LatentVelo~\cite{farrell2023inferring} incorporates latent variables $\boldsymbol{z}_c= (\boldsymbol{u}^{(z)}_c,\boldsymbol{s}^{(z)}_c)$ and pseudotime $t_{c}$, representing latent-space unspliced/spliced RNA levels and cellular pseudotime. It assumes the following dynamics:
$$
\begin{aligned}
        \dfrac{\mathrm{d}\boldsymbol{u}^{(z)}_c(t)}{\mathrm{d}t} &= f_u(\boldsymbol{u}^{(z)}_c(t),\boldsymbol{r}^{(z)}_c(t)), \\
        \dfrac{\mathrm{d}\boldsymbol{s}^{(z)}_c(t)}{\mathrm{d}t} &= f_s(\boldsymbol{u}^{(z)}_c(t),\boldsymbol{s}^{(z)}_c(t)), \\
        \dfrac{\mathrm{d}\boldsymbol{r}^{(z)}_c(t)}{\mathrm{d}t} &= f_r(\boldsymbol{s}^{(z)}_c(t),\boldsymbol{r}^{(z)}_c(t),\boldsymbol{h}_c), \\
        \boldsymbol{h}_c &= f_h(\boldsymbol{s}^{(z),\mathrm{obs}}_c, \boldsymbol{u}^{(z),\mathrm{obs}}_c),
\end{aligned}
$$
where $\boldsymbol{s}^{(z),\mathrm{obs}}_c$ and $\boldsymbol{u}^{(z),\mathrm{obs}}_c$ are latent-space observations, $\boldsymbol{r}^{(z)}_c$ governs chromatin dynamics, and~$f_u, f_s, f_r, f_h$ are neural networks with $f_h$ computing the cell state encoding $\boldsymbol{h}_c$. Beyond~standard VAE losses, an~\textbf{evolution loss} ensures correct dynamics:
$$
\mathcal{L}_{\mathrm{evol}} = \sum_{c=1}^{n_{c}} \mathbb{E}_{ t_c  \sim q(t_c|\boldsymbol{x})}\left[\dfrac{\|\boldsymbol{z}^{\mathrm{obs}}_c - \boldsymbol{z}_{c}(t_{c})\|^{2} + \|\boldsymbol{x}_{c} - \boldsymbol{\hat{x}}_{c}(t_{c})\|}{\sigma^2}\right].
$$
The total loss is $\mathcal{L} = \mathcal{L}_{\mathrm{rec}} + \mathcal{L}_{\mathrm{reg}} + \mathcal{L}_{\mathrm{evol}}$.

{VeloVI}~\cite{gayoso2024deep} models gene-specific state distributions $\boldsymbol{\pi}_{g,c}$ with a Dirichlet prior $\boldsymbol{\pi}_{gc} \sim \mathrm{Dirichlet}(\frac{1}{4},\frac{1}{4},\frac{1}{4},\frac{1}{4})$, where states $k_{g,c} \in \{1,2,3,4\}$ correspond to induction, repression, induction steady, and~repression steady. Key parameters include gene-specific rates $\alpha_{g}, \beta_{g}, \gamma_{g}$, pseudotime $t_{g,c}$, and~switching time $t_{g}^{s}$. For~repression-related states, $\alpha_{g}=0$ is fixed. Genes in transient states follow  Equation~(\ref{velocityeq}), while steady states use analytic solutions. Reconstruction losses follow standard VAE~training.

{VeloVAE}~\cite{gu2022variational} uses latent variables $\boldsymbol{z}_{c} \sim \mathcal{N}(0,I)$ and pseudotime $t_{c} \sim \mathcal{N}(t_{0}, \sigma_{0}^{2})$. A~fully connected network maps $\boldsymbol{z}$ to gene-specific parameters $\alpha_{c, g}, \beta_{c, g},\gamma_{c,g}$, enabling reconstruction via  Equation~(\ref{velocityeq}) and training with standard VAE~losses.

\paragraph{{Enhancing} Velocity: Continuity-Based Methods}

Several methods leverage the \textbf{continuity assumption} in single-cell data: If the observed data fully capture the continuous dynamics of cellular evolution, a~cell's state at the next timestep should align with its neighbors. Several methods use such prior information for defining the loss function to further refine the RNA velocity. 

{DeepVelo}~\cite{cui2024deepvelo} parameterizes $\alpha_{g,c}, \beta_{g,c}, \gamma_{g,c}$ via neural networks to compute RNA velocity $v_{g,c}$. Its loss function incorporates temporal consistency: The state $\boldsymbol{s}_c(t+1)$ is approximated as a weighted sum of neighboring cell states $\boldsymbol{s}_j(t(j))$, where the transition probability $\mathbb{P}_{+}(c \rightarrow j)$ is defined as
$$
\mathbb{P}_{+}(c \rightarrow j) = \begin{cases}
        \dfrac{1}{Z} & \text{if } \cos(\boldsymbol{s}_j - \boldsymbol{s}_c, \boldsymbol{\hat{v}}_c) > 0 \ \text{and} \ j \in \mathcal{N}(c), \\ 
        0 & \text{otherwise}.
    \end{cases}
$$
where $Z$ is the normalization constant. The~forward consistency loss enforces this assumption as follows:
$$
\mathcal{L}_{+}  = \sum_{c=1}^{n_{c}} \left\| \boldsymbol{s}_{c}(t) + \boldsymbol{\hat{v}}_{c} - \boldsymbol{s}_{c}(t+1) \right\|^2,
$$
with $\boldsymbol{s}_{c}(t+1) = \sum_{j \in \mathcal{N}(c)} \boldsymbol{s}_j(t(j)) \mathbb{P}_{+}(c \rightarrow j)$. A~symmetric backward consistency loss $\mathcal{L}_{-}$ is similarly constructed. Additionally, a~correlation loss ensures consistency with transcriptional dynamics in the following way:

$$
\mathcal{L}_{\mathrm{corr}} = - \left[ \lambda_{u} \dfrac{\boldsymbol{\hat{v}}_{c} \cdot \boldsymbol{u}_{c}}{\|\boldsymbol{\hat{v}}_{c}\| \|\boldsymbol{u}_{c}\|} + \lambda_{s} \dfrac{\boldsymbol{\hat{v}}_{c} \cdot \boldsymbol{s}_{c}}{\|\boldsymbol{\hat{v}}_{c}\| \|\boldsymbol{s}_{c}\|} \right],
$$
yielding the total loss $\mathcal{L} = \mathcal{L}_{+} + \mathcal{L}_{-} + \mathcal{L}_{\mathrm{corr}}$.

{CellDancer}~\cite{li2024relay} similarly parameterizes $\alpha_{g,c}, \beta_{g,c}, \gamma_{g,c}$ with neural networks, and~estimates velocity as\vspace{-6pt}
$$
\boldsymbol{\hat{v}}_c = (\Delta \boldsymbol{u}_c, \Delta \boldsymbol{s}_c), \quad \text{where } \Delta \boldsymbol{u}_c = \boldsymbol{\alpha}_c - \boldsymbol{\beta}_c \odot \boldsymbol{u}_c, \ \Delta \boldsymbol{s}_c = \boldsymbol{\beta}_c - \boldsymbol{\gamma}_c \odot \boldsymbol{s}_c.
$$
Its loss maximizes velocity alignment with neighbors, as~follows:\vspace{-6pt}
$$
\mathcal{L} = \sum_c \left[ 1 - \max_{j \in \mathcal{N}(c)} \cos\left(\boldsymbol{\hat{v}}_c, \boldsymbol{v}_j \right) \right], \quad \boldsymbol{v}_j = (\boldsymbol{u}_j - \boldsymbol{u}_c, \boldsymbol{s}_j - \boldsymbol{s}_c).
$$

{\textbf{Vector Field Reconstruction Based on RNA Velocity}}
After estimating the model parameters and obtaining the RNA velocity for each cell, an~important downstream task is to predict the fate of cells based on the estimated velocity, e.g.,~the continuous differentiation trajectory of cells. To~construct a continuous vector field, the~RNA velocity of each cell could be treated as the value of a vector field at discrete points. Then, the~vector field is reconstructed by formulating a regression problem. Subsequently, quantities derived from vector field analysis, such as equilibrium points, streamlines, gradients, divergence, and~curl, are used to study cell fate in continuous dynamics setup~\cite{qiu2022mapping}.

\paragraph{{Estimating} the Vector Field}

Denote the spliced RNA counts and RNA velocities at various data points $\{\boldsymbol{x}_i, \boldsymbol{v}_i\}_{i=1}^n$. The~task is to find a continuous vector field function $f^\star$ that minimizes the regression loss
$
    \mathcal{L} = \sum_i p_i \|\boldsymbol{v}_i - \boldsymbol{f}(\boldsymbol{x}_i)\|^2
$ where $p_i$ denotes the weights of each data~point.

Dynamo~\cite{qiu2022mapping} approximates the unknown vector-valued function in a sparse reproducing kernel Hilbert space (RKHS). For~a vector-valued function $\boldsymbol{f} \in \mathcal{H}$ in RKHS, it can be represented as a sum of Gaussian kernels as follows:\vspace{-6pt}
$$
    \boldsymbol{f}(\boldsymbol{x}) = \sum_{i=1}^m \Gamma(\boldsymbol{x}, \boldsymbol{\tilde{x}}_i)\boldsymbol{c}_i, \quad \Gamma(\boldsymbol{x}, \boldsymbol{\tilde{x}}) = \exp(- w \|\boldsymbol{x} - \boldsymbol{\tilde{x}}\|^2),
$$
where $\boldsymbol{\tilde{x}}_i$ are called control points. Additionally, the~norm of $\boldsymbol{f}$ in $\mathcal{H}$ can be computed as $ \|\boldsymbol{f}\|^2 = \sum_{i,j=1}^{m} \boldsymbol{c}_i^T \Gamma(\boldsymbol{\tilde{x}}_i, \boldsymbol{\tilde{x}}_j) \boldsymbol{c}_j^T.$
The loss function for the vector field estimation problem includes a regularization term based on the norm of $\boldsymbol{f}$ such that 
$
    \mathcal{L}_{\lambda} = \sum_i p_i \|\boldsymbol{v}_i - \boldsymbol{f}(\boldsymbol{x}_i)\|^2 + \frac{\lambda}{2} \|\boldsymbol{f}\|^2,
$
where $\lambda$ is the regularization~coefficient. 

Another popular fitting strategy to reconstruct the continuous vector field is to use the neural network, where a VAE-based deep learning method was proposed in~\cite{chen2022deepvelo}.

\paragraph{{Geometric} Analysis of Vector Field}

Based on the estimated continuous vector field, Dynamo~\cite{qiu2022mapping} proposed several analyses to reveal the differential geometry of the RNA velocity. First, the~Jacobian matrix is essential for analyzing the stability of equilibrium points in a dynamical system and for studying gene–gene interactions. In~RKHS context, the~Jacobian matrix can be analytically computed as
$$
    \boldsymbol{J} = \frac{\partial \boldsymbol{f}(\boldsymbol{x})}{\partial \boldsymbol{x}} = -2w \sum_{i=1}^m \Gamma(\boldsymbol{x}, \boldsymbol{\tilde{x}}_i) \boldsymbol{c}_i (\boldsymbol{x} - \boldsymbol{\tilde{x}}_i)^T
$$
where $\Gamma(\boldsymbol{x}, \boldsymbol{\tilde{x}}_i)$ is the Gaussian kernel. Since the $(i, j)$-th entry of the Jacobian matrix $J_{ij}$ represents the effect of unspliced RNA levels of gene $j$ on the RNA velocity of gene $i$, the~Jacobian matrix can be used to analyze the strength of gene–gene interactions. By~averaging the Jacobian matrix across all data points, an~average Jacobian matrix $\langle J \rangle$ can be obtained. By~sorting the elements in each row of $\langle J \rangle$, the~top regulators for each effector can be identified. Conversely, by~sorting the elements in each column of $\langle J \rangle$, the~top effectors for each regulator can be identified. The~Jacobian matrix can also be used to compute the effect of perturbations. If~the system state changes by $\Delta \boldsymbol{x}$ at $\boldsymbol{x}$, the~resulting change in the vector field is
$
    \Delta \boldsymbol{f} = \boldsymbol{J} \Delta \boldsymbol{x}.
$

\textls[-15]{Several other quantities could also be conveniently derived based on the Jacobian~matrix.}
\begin{itemize}
    \item The divergence represents the net flux generated or dissipated per unit time at each point in the vector field:
$$
    \nabla \cdot \boldsymbol{f} = \mathrm{Tr}(\boldsymbol{J}),
$$
where $\mathrm{Tr}(J)$ denotes the trace of the Jacobian matrix. Regions with divergence greater than 0 (sources) may correspond to the initial states of cells, while regions with divergence less than 0 (sinks) may correspond to the terminal states of~cells.

\item The acceleration of a particle moving along the streamlines of the vector field can be directly computed from the Jacobian matrix:\vspace{-6pt}
$$
    \boldsymbol{a} = \frac{\mathrm{d} \boldsymbol{v}}{\mathrm{d}t} = \boldsymbol{J} \boldsymbol{v}.
$$

\item The curvature vector of the streamlines is defined as the derivative of the unit tangent vector with respect to time:\vspace{-6pt}
$$
    \boldsymbol{\kappa} = \frac{1}{\|\boldsymbol{v}\|} \frac{\mathrm{d}}{\mathrm{d}t} \frac{\boldsymbol{v}}{\|\boldsymbol{v}\|} = \frac{ (\boldsymbol{v}^T  \boldsymbol{v})\boldsymbol{J} \boldsymbol{v} -  (\boldsymbol{v}^T \boldsymbol{J} \boldsymbol{v})\boldsymbol{v}}{\|\boldsymbol{v}\|^4}.
$$
\end{itemize}

\paragraph{{Transition} Path Analysis}

Based on the learned vector field, continuous cell trajectories could also be constructed in Dynamo~\cite{qiu2022mapping} using the concept of most probable path~\cite{landscape_PZ}. Given the SDE \eqref{eq:sde}, the~action along  any path $\Psi$ is defined as
$$
S_{T}[\psi] = \int_{0}^{T} L^{FW}\left(\psi,\dot{\psi}\right) dt,L^{FW}\left(\psi,\dot{\psi}\right) = \frac{1}{4}\left[\dot{\psi}(s)-b\left(\psi(s)\right)\right]^{t} D^{-1}\left(\psi(s)\right)\left[\dot{\psi}(s)-b\left(\psi(s)\right).\right]
$$
According to the Freidlin–Wentzell theorem~\cite{landscape_PZ}, the~path of least action is indeed the most probable path to make transitions between two attractors. In~actual computation, Dynamo assumes the constant noise coefficient by taking $D = \frac{\sigma^2}{2}$ since only a continuous vector field is~reconstructed.

For a transition connecting two meta-stable states along the optimal path with action $S^\star \geq 0$, the~transition rate between the attractors is given by\vspace{-6pt}
\begin{equation}\label{eq:transitionrate}
        R(\boldsymbol{x}_s \rightarrow \boldsymbol{x}_t) \approx C \exp(-S^\star),
\end{equation}
where $C$ is a proportionality~constant.

\subsection{Temporally Resolved Single-Cell~RNA-Seq}

Temporally resolved scRNA-seq provides us with a deeper understanding of the dynamics process in single cells.  However, due to the destructive nature of scRNA-seq technology, we cannot track the trajectories of individual cells. Instead, we can only observe the changes in cellular distribution with time. Thus, reconstructing the trajectories of single cells from samples collected at discrete and sparse temporal points becomes crucial for understanding developmental processes and other dynamic biological processes and remains a challenging problem~\cite{waddingot,Tigon,peng2024stvcr,trajectory,jiang2022dynamic,jiang2024physics,bunne2024optimal,bunne2023learning,tong2023unblanced,Stephenzhang2021optimal,maddu2024inferring,eyring2024unbalancedness,scnode,PRESCIENT}. To~overcome these challenges, many methods have been developed in recent years. From~a dynamical perspective, these approaches can be broadly classified into two categories: those that model dynamics on discrete cell states and those that model dynamics in continuous spaces. In~the following, we will introduce these methods separately from these two viewpoints. {Figure \ref{fig:sec2_2_temporal} summarizes these approaches.}

\vspace{-3pt}
\begin{figure}[H]

\begin{adjustwidth}{-\extralength}{0cm}
\centering 
        \includegraphics[width=0.95\linewidth]{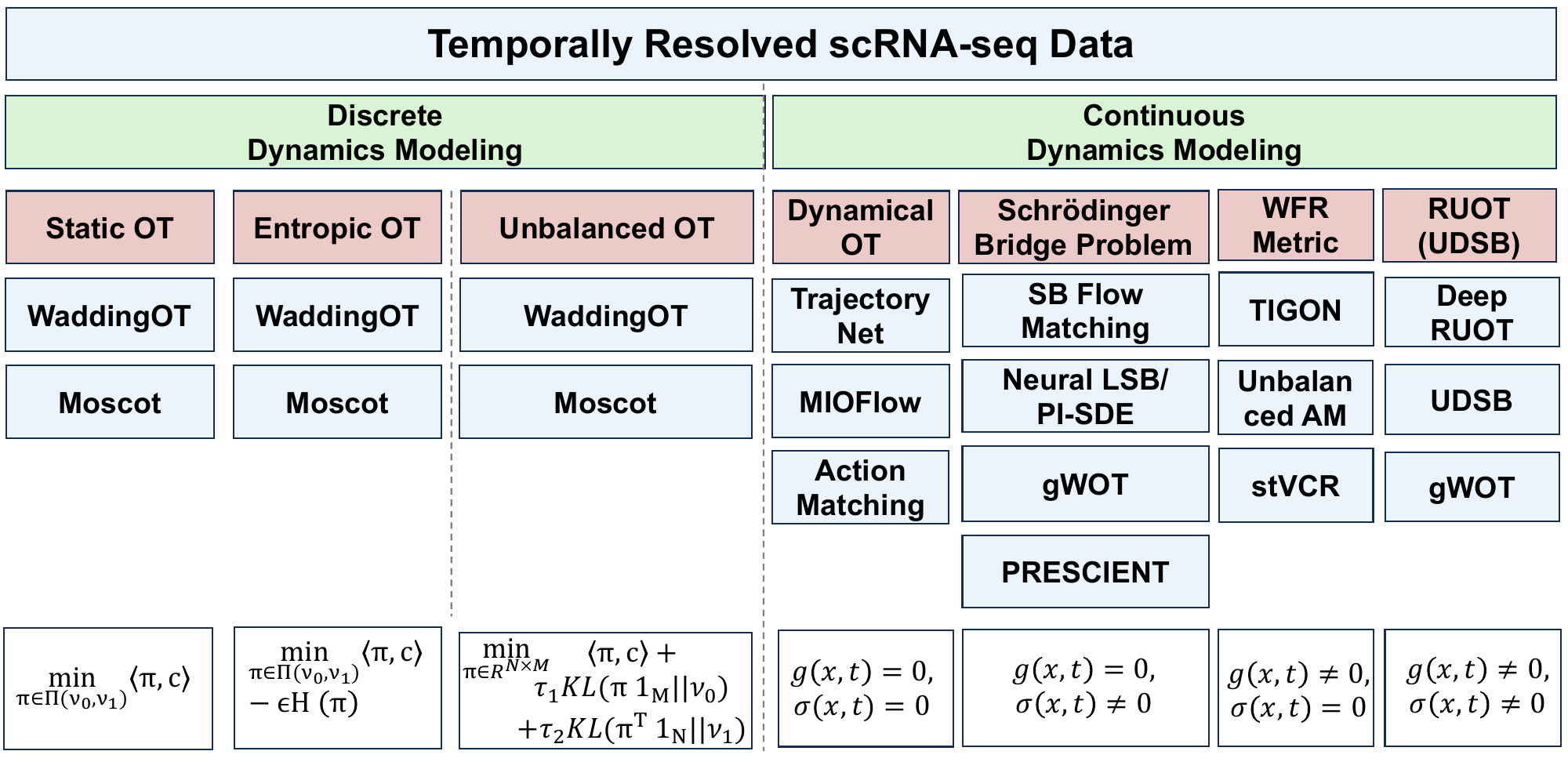}
\end{adjustwidth}
        \caption{\textbf{{Dynamic} 
 modeling of temporally-resolved single-cell transcriptomics.}}
        \label{fig:sec2_2_temporal}

\end{figure}
\subsubsection{Discrete Temporal Dynamics~Modeling}
Among the methods that model dynamics on discrete cell states, pioneering work includes Waddington OT~\cite{waddingot} and Moscot~\cite{moscot}. These approaches employ static optimal transport as the main~tools. 

{\textbf{Static Optimal Transport}}
To formulate this problem,  consider $\boldsymbol{X} \in \mathbb{R}^{N \times G}$ and $\boldsymbol{Y}\in \mathbb{R}^{M \times G}$ represent two unpaired datasets of $N$ and $M$ cells observed at different time points ($t_1, t_2$), respectively, in~the $G$ dimensional gene expression space. Then, one can define two marginal distributions  $\nu_0 \in \mathcal{C}_N$, $\nu_1 \in \mathcal{C}_M$ at $t_1$ and $t_2$ respectively  on  the probability simplex $\mathcal{C}_N=\{\boldsymbol{a} \in \mathbb{R}^N |\sum_{i=1}^N a_i=1, \boldsymbol{a} \geq 0\}.$
The goal of optimal transport is to find the optimal coupling $\boldsymbol{\pi} \in \mathbb{R}_{+}^{N \times M}$ that transports a distribution to another while minimizing the cost associated with the transportation. The~feasible transport plan is defined as 
$\boldsymbol{\boldsymbol{\Pi}}(\nu_0, \nu_1)=\left\{\boldsymbol{\pi} \in \mathbb{R}^{N \times M}: \boldsymbol{\pi} \boldsymbol{1}_M=\nu_0, \boldsymbol{\pi}^T \boldsymbol{1}_N=\nu_1, \boldsymbol{\pi} \geq 0\right\}.$
So the static optimal transport problem is formally defined as\vspace{-6pt}
\begin{equation}
    \begin{aligned}\label{eq:static_OT}
     \min_{\boldsymbol{\pi} \in \boldsymbol{\Pi}(\nu_0, \nu_1)}\langle\boldsymbol{\pi}, \boldsymbol{c}\rangle=\sum_{i, j}c_{i, j}\pi_{i, j}.
\end{aligned}
\end{equation}
The cost matrix \(\boldsymbol{c} \in \mathbb{R}_{+}^{n \times m}\) defines the transportation cost between each pair of points, where \(c_{ij}:= c(\boldsymbol{x}_i, \boldsymbol{y}_j)\) quantifies the expense of transferring a unit mass from the source point  $\boldsymbol{x}_i$  to the target point  $\boldsymbol{y}_j$. By~solving the static optimal transport problem, one can determine the transport matrix that couples the two distributions. This static optimal transport can be effectively addressed using the Python Optimal Transport (POT) library~\cite{flamary2021pot}. 

{\bf \textbf{Entropic Optimal Transport}}
Additionally, to~enhance the efficiency of solving the optimal transport problem, a~regularized optimal transport approach is often introduced. The~discrete entropy of a coupling matrix is defined~as
$$
H(\boldsymbol{\pi}) \stackrel{\text { def }}{=}-\sum_{i, j} \pi_{i, j}\left(\log \left(\pi_{i, j}\right)-1\right).
$$
The function $H$ is 1 -strongly concave, because~its Hessian is $\partial^2 H(\boldsymbol{\pi})=-\operatorname{diag}\left(1 / \pi_{i, j}\right)$ and $0<\pi_{i, j} \leq 1$. The~idea of the entropic regularization of optimal transport is to use $-H$ as a regularizing function to obtain approximate solutions to the original transport problem:\vspace{-6pt}
\begin{equation}\label{eq:ethropicot}
    \min _{\boldsymbol{\pi} \in \boldsymbol{\Pi}(\boldsymbol{\nu_0}, \boldsymbol{\nu_1})}\langle\boldsymbol{\pi}, \boldsymbol{c}\rangle-\varepsilon H(\boldsymbol{\pi}) 
\end{equation}
Since the objective is an $\varepsilon$-strongly convex function, problem \eqref{eq:ethropicot} has a unique optimal solution. Using the KKT conditions,
the solution to \eqref{eq:ethropicot} is unique and has the form~\cite{ot_theory}\vspace{-6pt}
$$
\forall(i, j) \in(N \times M), \quad \pi_{i, j}=a_i K_{i, j} b_j
$$
for two unknown variables $(\boldsymbol{a}, \boldsymbol{b}) \in \mathbb{R}_{+}^N \times \mathbb{R}_{+}^M$, where $K_{i, j}=e^{-\frac{c_{i, j}}{\varepsilon}}$.

{\emph{Sinkhorn Algorithm}} From above,  
 the optimal solution of problem \eqref{eq:ethropicot} can be  expressed in matrix form as $\boldsymbol{\pi}=\operatorname{diag}(\boldsymbol{a}) \boldsymbol{K} \operatorname{diag}(\boldsymbol{b})$. Then it is necessary to satisfy the constraints $\boldsymbol{\Pi}(\nu_0, \nu_1)$, i.e.,\vspace{-6pt}
$$
\operatorname{diag}(\boldsymbol{a}) \boldsymbol{K} \operatorname{diag}(\boldsymbol{b})\boldsymbol{1}_M=\nu_0, \quad \operatorname{diag}(\boldsymbol{a}) \boldsymbol{K}^{T} \operatorname{diag}(\boldsymbol{b})\boldsymbol{1}_N=\nu_1 .
$$
Note that $\operatorname{diag}(\boldsymbol{b}) \boldsymbol{1}_M$ is $\boldsymbol{b}$ and 
 the multiplication of $\operatorname{diag}(\boldsymbol{a})$ times $\boldsymbol{K} \boldsymbol{b}$ is
$$
\boldsymbol{a}\odot(\boldsymbol{K} \boldsymbol{b})=\nu_0,  \boldsymbol{a} \odot\left(\boldsymbol{K}^{\mathrm{T}} \boldsymbol{b}\right)=\nu_1,
$$
where $\odot$ represents the entry-wise multiplication. An~intuitive way is to solve them iteratively, and~these two updates yield the Sinkhorn algorithm,
$$
\boldsymbol{a}^{(\ell+1)} \stackrel{\text { def. }}{=} \frac{\nu_0}{\boldsymbol{K} \boldsymbol{b}^{(\ell)}}, \quad  \quad \boldsymbol{b}^{(\ell+1)} \stackrel{\text { def. }}{=} \frac{\nu_1}{\boldsymbol{K}^{{T}} \boldsymbol{a}^{(\ell+1)}}.
$$
 The division used above between two vectors is entry-wise and can be computed in time and memory quadratically of cell~number.

\textbf{Unbalanced Optimal Transport}
Static optimal transport inherently conserves mass. However, during~cellular development and differentiation, processes such as cell proliferation and apoptosis result in mass non-conservation. Consequently, it is essential to consider unnormalized distributions that account for cell growth and death. Additionally, the~marginals must be adjusted to incorporate these factors effectively. So, the~unbalanced optimal transport can be defined as follows:
\begin{equation} \label{eq:static_ubot}
 \min_{\boldsymbol{\pi} \in \mathbb{R}_{+}^{N \times M}}\langle\boldsymbol{\pi}, \boldsymbol{c}\rangle+\tau_1 \operatorname{KL}(\boldsymbol{\pi} \boldsymbol{1}_M||\nu_0)+\tau_2 \operatorname{KL}(\boldsymbol{\pi}^T \boldsymbol{1}_N||\nu_1),
\end{equation}
where $\tau_1$ and $\tau_2$ are hyperparameters that control the degree of penalization. When $\tau_1=\tau_2 \rightarrow +\infty$, then one can recover the original optimal transport. To~further adjust the marginal distributions accounting for growth and death~\cite{moscot,waddingot}, for~the left marginal distribution $\nu_0$ we set\vspace{-6pt}
\begin{equation} \label{eq:static_ubot_v0}
(\nu_0)_i=\frac{g\left(\boldsymbol{x}_i\right)^{t_2-t_1}}{\sum_{j=1}^N g\left(\boldsymbol{x}_j\right)^{t_2-t_1}}, \quad \forall i \in\{1, \ldots, N\}.
\end{equation}
where $g$ is the growth/death function and can be estimated through the gene sets. For~the right marginal distribution $\nu_1$, set it as the uniform distribution, i.e.,~$(\nu_1)_j = 1/M, \linebreak \quad \forall j \in\{1, \ldots, M\}.$

 By the obtained coupling matrix $\boldsymbol{\pi} \in \mathbb{R}_{+}^{N \times M}$, one can perform biological downstream analysis such as computing ancestors or descends of a cell state and imputing gene expressions~\cite{waddingot,moscot}. Naturally, a~Markov chain model can be formulated to quantify the transition probability among cells across time points based on the optimal transport plan~\cite{moscot,weiler2024cellrank}. By~weighting this random walk with those induced by other quantities such as gene expression similairity (e.g., DiffuionMap), pseudotime (e.g., Palantir), or~RNA velocity, the~CellRank analysis could also be conducted to dissect the underlying structure of the transitional dynamics~\cite{weiler2024cellrank}.

\subsubsection{Continuous Temporal  Dynamics~Modeling}
Although static optimal transport provides a robust framework for coupling distributions at different time points, there is a substantial interest in capturing continuous cellular dynamics over time and fitting mechanistic models that transform the source distribution into the target distribution. This interest has driven the development of various dynamical optimal transport (OT) methods and flow-based generative models. Prominent approaches include those based on the Benamou–Brenier formulation~\cite{benamou2000computational}, such as TrajectoryNet~\cite{trajectorynet}, MIOFlow~\cite{mioflow}, and~other related methodologies~\cite{Liwu2020machine, zhouhaolearning, shenjie2024new, DOngbin2023scalable, pooladian2024neural, scnode}. Additionally, unbalanced dynamic OT methods~\cite{peng2024stvcr, Tigon, tong2023unblanced, eyring2024unbalancedness}, Gromov–Wasserstein OT approaches~\cite{genot}, continuous normalizing flows (CNF), and~conditional flow matching techniques (CFM)~\cite{cfm_lipman, cfm_tong, albergo2023building, liu2023flow, jiao2024convergenceanalysisflowmatching, jiao2024convergencecontinuousnormalizingflows, zhouhao2022neural, zhouhao2024parameterized_H, zhouhao2024parameterized_W, LiwuZHou2020wasserstein, Lujianfeng2024convergence} have also been proposed. Despite these advancements, many of these methods do not fully account for stochastic dynamical effects, particularly the intrinsic noise inherent in gene expression and cell differentiation~\cite{zhou2021stochasticity, zhou2021dissecting}, which are prevalent in single-cell biological processes~\cite{elowitz2002stochastic}.

In the realm of stochastic dynamics, the~Schrödinger bridge (SB) problem seeks to identify the most probable stochastic transition path between two arbitrary distributions relative to a reference stochastic process~\cite{sb}. Variants of the SB problem have been applied across various domains, including single-cell RNA sequencing (scRNA-seq) analysis and generative modeling. These approaches encompass static methods~\cite{trajectory, bunne_unsb, Geofftrajectory, Geofftrajectory2, shi2024diffusion, de2021diffusion, pooladian2024plug, Liudeep_genera_sb,gu2024partially}, dynamic methods~\cite{dyn_sb_koshizuka2023neural, action_matching, tong_action, Duan_action, bunne_dynam_SB, chen2022likelihood, albergo2023stochastic, wang2021deep, Jiao2024LatentSB, zhou2024denoising, imagesb, LIWUCHEN_score,Linwei2024governing,maddu2024inferring,PRESCIENT,jiang2024physics}, and~flow-matching techniques~\cite{sflowmatch}. However, these methods often fail to address unnormalized distributions resulting from cell growth and death. To~address this problem, some methods have been developed to account for the unbalanced stochastic dynamics, for~example, those based on branching SDE theory (e.g, \mbox{gWOT)~\cite{Geofftrajectory2,Stephenzhang2021optimal,Geofftrajectory,trajectory}} and those based on the Feynman–Kac formula with forward-backward SDE theory~\cite{bunne_unsb}. Among~those, most methods often require prior knowledge of these processes, such as growth or death rates ~\cite{bunne_unsb, trajectory,Geofftrajectory2,Stephenzhang2021optimal}
 or depend on additional information like cell lineage data~\cite{Geofftrajectory}.
 
Recently, regularized unbalanced optimal transport (RUOT), also known as unbalanced  Schrödinger bridge~\cite{chen2022most}, has emerged as a promising approach for modeling stochastic unbalanced continuous dynamics~\cite{branRUOT, chen2022most, highorderRUOT, janati2020entropic}. RUOT can be viewed as an unbalanced relaxation of the dynamic Schrödinger bridge formulation. For~instance,~{ref.}~\cite{branRUOT} elucidates the connection between certain RUOT formulations and branching Schrödinger bridges. Meanwhile, a~new deep learning framework (DeepRUOT)~\cite{DeepRUOT}, has been developed to learn general RUOT and infer continuous unbalanced stochastic dynamics from sample data based on derived Fisher regularization forms without requiring prior~knowledge.

\textls[-5]{The primary objective now transforms to determine the dynamics described by \mbox{Equations} \eqref{eq:sde} and~(\ref{eq:pde})} from observed data, given unnormalized distributions at  $T$  discrete time points where \(\boldsymbol{x}_i \in \mathbb{R}^d \sim {\nu}_i \) for each fixed time point  $i \in \{0, \ldots, T-1\}$. Note that solely satisfying Equation~(\ref{eq:pde}) does not admit a unique solution. Consequently, we need to ensure that the inferred dynamics also adhere to certain energy minimization principles. Building upon Equation~(\ref{eq:pde}), the~problem can be categorized into four distinct scenarios:\linebreak 
	(1)	 $\gxt = 0  \text{ and }  \dxt = 0,$
	(2)	 $\gxt = 0   \text{ and }  \dxt \neq 0,$
	(3)	$ \gxt \neq 0   \text{ and }  \linebreak \dxt = 0,$
	(4)	$ \gxt \neq 0  \text{ and } \dxt \neq 0 .$
Each of these cases is examined to systematically address the learning of the underlying~dynamics.

{\textbf{Dynamical Optimal Transport ($g = 0, \sigma = 0$)}} In this case, it means the dynamics do not account for unblancedness and stochastic, i.e., the~cellular dynamics are governed by $\rmd \xt=\boldsymbol{b}(\xt, t)\rmd t$. Then we can use the dynamical optimal transport to model these dynamics, also known as the Benamou–Brenier formulation~\cite{benamou2000computational}, which can be stated as~follows:
\begin{equation}\label{eq:dynamicalot}
\begin{gathered}
\dfrac{1}{2}\mathcal{W}_2^2\left(\nu_0, \nu_1\right)=\inf _{\left(\pxt, \bxt\right)} \int_0^1 \int_{\mathbb{R}^d} \frac{1}{2}\|\bxt\|_2^2 \pxt\rmd \boldsymbol{x} \rmd t, \\
\mbox{s.t. }\partial_t p+\nabla \cdot\left(\bxt p\right)=0,\ p|_{t=0}=\nu_0, p|_{t=1}=\nu_1 .
\end{gathered}
\end{equation}
The inclusion of the factor $\frac{1}{2}$ on the left-hand side ensures that the Wasserstein distance has a more physically meaningful interpretation; for example, it represents the total action required to transport one distribution to another. In~this formulation, probability distributions are connected through a deterministic transport equation. It has been demonstrated that this dynamic formulation is equivalent to static optimal transport problem \linebreak {(Equation \eqref{eq:static_OT})} 
 when employing the cost function  $c(\boldsymbol{x}, \boldsymbol{y}) = \|\boldsymbol{x} - \boldsymbol{y}\|_2^2$ . 

\paragraph{{Neural} ODE Solver} 

Numerous methodologies have been proposed to solve dynamical optimal transport or its variants numerically. The~basic approach involves employing a neural network, denoted as  $\boldsymbol{b}_\theta (\boldsymbol{x}, t)$, to~parameterize  $\bxt$, and~subsequently utilizing the ordinary differential equations (ODEs) that govern particle trajectories. From~Problem \eqref{eq:dynamicalot}, it is evident that the optimization process must address two distinct loss components: the first pertains to the computation of an energy-related loss, while the second concerns the reconstruction error ($p (\boldsymbol{x}, 1)= \nu_1$).

Regarding energy loss, the~high-dimensional nature of the integral presents significant challenges due to the curse of dimensionality. To~mitigate this issue, the~integral is approximated using Monte Carlo integration and continuous normalizing flow (CNF) techniques~\cite{trajectorynet}. The~strategy involves performing integration along the particle trajectories dictated by the ODE, i.e.,
$$
 \int_0^1 \int_{\mathbb{R}^d} \frac{1}{2}\|\bxt\|_2^2 \pxt\rmd \boldsymbol{x} \rmd t = \mathbb{E}_{\boldsymbol{x}_0 \sim \nu_0}\int_{0}^{1}\frac{1}{2}\|\boldsymbol{b} (\boldsymbol{x}(t), t)\|_2^2 \rmd t,
$$
where $\boldsymbol{x}(t)$ satisfies the ODE $\frac{\rmd \boldsymbol{x}}{\rmd t}=\bxt, \boldsymbol{x}_0 \sim \nu_0$. For~the distribution reconstruction loss, the~authors incorporate an additional penalizing constraint. By~integrating these two loss components, 
   there exists a sufficiently large $\lambda \geq 0$ such that~\cite{trajectorynet,mioflow}
    $$
 \dfrac{1}{2}\mathcal{W}_2^2\left(\nu_0, \nu_1\right)=\inf_{(\pxt, \bxt)}\mathbb{E}_{\boldsymbol{x}_0 \sim \nu_0}\int_{0}^{1}\frac{1}{2}\|\boldsymbol{b} (\boldsymbol{x}(t), t)\|_2^2 \rmd t + \lambda \mathcal{D}(p (\boldsymbol{x},1), \nu_1).
    $$
Based on this formulation, TrajectoryNet~\cite{trajectorynet}  computes both the energy and the reconstruction errors by using neural ODE~\cite{chen2018neuralode} to parametrize the velocity $\bxt$.

\paragraph{{Conditional} Flow Matching} 

Recently, conditional flow matching (CFM) presents another efficient dynamical OT solver especially in high dimensionality case~\cite{cfm_lipman,cfm_tong,liu2023flow,albergo2023stochastic}. Assume that the probability path  $\pxt$  and the corresponding vector field  $\bxt$  generating it are known, and~that  $\pxt$  can be efficiently sampled. Under~these conditions, a~neural network $\boldsymbol{b}_\theta (\boldsymbol{x}, t)$  can be trained to approximate  $\bxt$  by minimizing the flow matching (FM) objective: 
$$
\mathcal{L}_{\text{FM}}(\theta)=\mathbb{E}_{t \sim \mathcal{U}(0,1), \boldsymbol{x} \sim \pxt }\|\boldsymbol{b}_\theta (\boldsymbol{x}, t)-\bxt\|_2^2.
$$
However, this objective is computationally intractable when dealing with general source and target distributions. Consider the specific case of  Gaussian marginal densities, defined as  $\pxt = \mathcal{N}(\boldsymbol{x} |\boldsymbol{\mu}(t), \sigma(t)^2\boldsymbol{I})$. The~corresponding unique vector field that generates this density from  $ \mathcal{N}(\boldsymbol{x} |\boldsymbol{\mu}(0), \sigma(0)^2\boldsymbol{I})$
 is $
\bxt=\boldsymbol{\mu}'(t)+\frac{\sigma(t)}{\sigma'(t)}(\boldsymbol{x}-\boldsymbol{\mu}(t)),
$
where $\mu'(t)$ and $\sigma'(t)$ means the time derivative~\cite{cfm_tong,cfm_lipman}. Now, assume the marginal probability trajectory  $\pxt$  is a mixture of conditional probability paths $p (\boldsymbol{x}, t |\boldsymbol{z})$. Specifically, this can be expressed as:
$$
\pxt =\int p (\boldsymbol{x}, t |\boldsymbol{z}) q(\boldsymbol{z}) \rmd \boldsymbol{z}.
$$
If the $ p (\boldsymbol{x}, t |\boldsymbol{z})$ is generated by the vector field $ \boldsymbol{b} (\boldsymbol{x}, t |\boldsymbol{z})$ from $ p (\boldsymbol{x}, 0 |\boldsymbol{z})$, then $\pxt$ can be generated by $\bxt$ defined as follows:\vspace{-6pt}
$$
\bxt:=\mathbb{E}_q(\boldsymbol{z})\frac{\boldsymbol{b} (\boldsymbol{x}, t |\boldsymbol{z})p (\boldsymbol{x}, 0 |\boldsymbol{z})}{\pxt}.
$$
This is also intractable since $\pxt$ is difficult to compute. The~key is to introduce the conditional flow matching objective:
$$
\mathcal{L}_{\text{CFM}}(\theta)=\mathbb{E}_{t \sim \mathcal{U}(0,1), q(\boldsymbol{z}),  p (\boldsymbol{x}, t |\boldsymbol{z})}\|\boldsymbol{b}_\theta (\boldsymbol{x}, t)-\boldsymbol{b} (\boldsymbol{x}, t |\boldsymbol{z})\|_2^2.
$$
One can prove that $\nabla_\theta \mathcal{L}_{\text{CFM}}=\nabla_\theta \mathcal{L}_{\text{FM}}$, so training with CFM is equivalent with FM. The~CFM objective is very useful when the  $\bxt$ is intractable but the conditional  $\boldsymbol{b} (\boldsymbol{x}, t |\boldsymbol{z})$ is tractable. So to approximate the dynamical optimal transport \eqref{eq:dynamicalot} is to use CFM. Assume $q(\boldsymbol{z})=q(\boldsymbol{z}_0, \boldsymbol{z}_1)$, and~set $q(\boldsymbol{z})$ to be the Wasserstein optimal transport map $\boldsymbol{\pi}$ between the source distribution $\nu_0$ and the target distribution $\nu_1$, i.e.,~$q(\boldsymbol{z})=\boldsymbol{\pi}(\boldsymbol{z}_0, \boldsymbol{z}_1)$, where $\boldsymbol{z}_0 \sim \nu_0$, $\boldsymbol{z}_1 \sim \nu_1$. 
Then one can construct the Gaussian flow between $\boldsymbol{z}_0$ and $\boldsymbol{z}_1$ with standard deviation $\sigma$,
$$
p (\boldsymbol{x}, t |\boldsymbol{z})=\mathcal{N} (\boldsymbol{x}|t \boldsymbol{z}_1 + (1-t) \boldsymbol{z}_0| \sigma^2),
\boldsymbol{b} (\boldsymbol{x}, t |\boldsymbol{z})=\boldsymbol{z}_1-\boldsymbol{z}_0.
$$
It can be proved that when $\sigma \rightarrow 0$,  this also gives a way to solve the dynamical optimal transport~\cite{cfm_tong}. The~advantage of CFM is that it is simulation-free and can handle the thousand gene dimensions without reducing~dimensionality.

{\bf \textbf{Schrödinger Bridge Problem ($g = 0,  \sigma \neq 0$)}} In this case, the~model can account for the stochastic 
effects, yet without unbalanced effects.  We employ the Schrödinger Bridge problem to model the SDE dynamics, i.e.,  the~cellular dynamics are \linebreak $\mathrm{d} \xt=\boldsymbol{b}\left(\xt, t\right) \mathrm{d} t+\boldsymbol{\sigma}\left(\xt, t\right)\mathrm{d}\boldsymbol{w}_t$. The~Schrödinger Bridge problem seeks to determine the most probable evolution between a specified initial distribution $\nu_0$ and a terminal distribution $\nu_1$ (assumed to possess a density in this study) relative to a given reference stochastic process. Formally, this \mbox{problem} is \mbox{formulated} as the minimization of the Kullback–Leibler (KL) divergence from the perspective of optimal control~\cite{dai1991stochastic}, as~shown below:
~\vspace{-6pt}
\begin{equation}
\label{eq:sb}
\min_{\mu^\Xb_0=\nu_0, \mu^\Xb_1=\nu_1} \mathcal{D}_{\mathrm{KL}}\left(\mu^\Xb_{[0,1]} | \mu^\Yb_{[0,1]}\right),
\end{equation}
where \(\mu^\Xb_{[0,1]}\) denotes the probability measure induced by the stochastic process \(\xt\) for $0 \leq t \leq 1$, defined on the space of all continuous paths $C([0,1], \mathbb{R}^d)$. The~distribution of \(\xt\) at a given time t is characterized by the measure \(\mu^\Xb_{t}\) with density function \(p(\boldsymbol{x}, t)\). The~reference measure \(\mu^\Yb_{[0,1]}\) is chosen as the probability measure induced by the process
\(\rmd \boldsymbol{Y}_t = \boldsymbol{\sigma}(\boldsymbol{Y}_t, t) \rmd \boldsymbol{w}_t,
\)
where \(\boldsymbol{w}_t \in \mathbb{R}^d\) represents the standard multidimensional Brownian~motion.

Interestingly, the~problem can be equivalently transformed into a dynamical form \cite{sb_Gentil, sb_chen, dai1991stochastic,DeepRUOT}
\begin{equation}
\label{eq:dsb2}
\inf_{\left(p, \boldsymbol{b}\right)} \int_0^1 \int_{\mathbb{R}^d} \left[ \frac{1}{2} \boldsymbol{b}^\top(\boldsymbol{x}, t) \axt^{-1} \boldsymbol{b}(\boldsymbol{x}, t) \right] p(\boldsymbol{x}, t) \rmd \boldsymbol{x} \rmd t,
\end{equation}
\textls[-25]{where the infimum is taken over all pairs of functions $(p, \boldsymbol{b})$ satisfying $p(\cdot, 0) = \nu_0$, $p(\cdot, 1) = \nu_1$,} and~$p(\boldsymbol{x}, t)$ is absolutely continuous with respect to time. Additionally, the~pair $(p, \boldsymbol{b})$ must satisfy the Fokker–Planck Equation \eqref{eq:pde}. We denote minimization problem \eqref{eq:dsb2} and the constraints \eqref{eq:pde} as the \emph{dynamic diffusion Schrödinger bridge} formulation.
Methods for modeling stochastic dynamics based on it have been widely \mbox{developed~\cite{maddu2024inferring,LIWUCHEN_score,jiang2024physics,PRESCIENT,dyn_sb_koshizuka2023neural,sflowmatch},} involving neural  SDE, neural ODE, or~flow matching techniques. We will next provide an overview of the methodologies in these~approaches. 

\paragraph{{Neural} SDE Solver} 

Similar to dynamical OT, one can solve the dynamical SB problem through the CNF formulation
    $$
\inf_{(\pxt, \bxt)}\mathbb{E}_{\boldsymbol{x}_0 \sim \nu_0}\int_{0}^{1}\left[ \frac{1}{2} \boldsymbol{b}^\top(\boldsymbol{x}(t), t) \boldsymbol{a}(\boldsymbol{x}(t),t)^{-1} \boldsymbol{b}(\boldsymbol{x}(t), t) \right] \rmd t + \lambda \mathcal{D}(p (\boldsymbol{x},1), \nu_1).
    $$
Building upon this, one can parametrize $\bxt$ and $\dxt$
 using neural networks respectively and solve this formulation through POT and the neural SDE solver. However, besides~these two terms, some work also introduces the idea of the principle of least action along the trajectory in which the optimal path has the smallest action value~\cite{jiang2024physics,dyn_sb_koshizuka2023neural}. Thus, they introduce a new Hamilton–Jacobi–Bellman (HJB) regularization term~\cite{jiang2024physics} when assuming $\bxt=-\nabla_{\boldsymbol{x}}\Phi (\boldsymbol{x}, t)$, i.e.,
 $$
 \mathcal{R}_h=\int_{0}^1 \int_{\mathbb{R}^d} \left|\partial_t \Phi (\boldsymbol{x}, t) -\|\nabla_{\boldsymbol{x}}\Phi (\boldsymbol{x}, t)\|_2^2\right|\pxt \rmd \boldsymbol{x} \rmd t.
 $$
 or a general form derived in~\cite{dyn_sb_koshizuka2023neural}. 
 
\paragraph{{Shr\"{o}dinger} Bridge Conditional Flow Matching} 

By leveraging CFM techniques, the~simulation-free Shr\"{o}dinger bridge~\cite{sflowmatch} has also been recently developed. The~core idea is to decompose the problem into a sequence of elementary conditional subproblems, each of which is more tractable, and~subsequently express the overall solution as a mixture of the solutions to these conditional subproblems. Let the reference process be a Brownian motion (i.e., $\boldsymbol{Y} = \sigma \boldsymbol{W}$). In~this case, the~Schrödinger bridge problem admits a unique solution $\mathbb{P}^*$, which is expressed as a mixture of Brownian bridges weighted by an entropic optimal transport (OT) plan:
\begin{equation}\label{eq:brownianbridge}
\mathbb{P}^*\left(\left(\boldsymbol{x}_t\right)_{t \in [0,1]}\right) = \int \mathbb{W}\left(\left(\boldsymbol{x}_t\right) \mid \boldsymbol{x}_0, \boldsymbol{x}_1\right) \, \rmd \pi_{2 \sigma^2}^{\star}\left(x_0, x_1\right),\end{equation}
where $\mathbb{W} \left(\left(\boldsymbol{x}_t\right)\mid{t \in (0,1)} \mid \boldsymbol{x}_0, \boldsymbol{x}_1 \right)$ denotes the Brownian bridge between $\boldsymbol{x}_0$ and $\boldsymbol{x}_1$ with a diffusion rate $\sigma$, and~$\pi_{2\sigma^2}^{\star}(\boldsymbol{x}_0, \boldsymbol{x}_1)$ represents the entropic optimal transport plan between the distributions. The~calculation of $\mathbb{W} \left(\left(\boldsymbol{x}_t\right)\mid{t \in (0,1)} \mid \boldsymbol{x}_0, \boldsymbol{x}_1 \right)$ can be framed as an optimal control problem:
\begin{equation*}
    \begin{aligned}
& \min_{\boldsymbol{b}} \mathbb{E}\int_0^1 \left\| \boldsymbol{b}\left(\xt, t\right) \right\|^2 \, dt,  \\
&   \mathrm{d} \xt=\boldsymbol{b}\left(\xt, t\right) \mathrm{d} t+\boldsymbol{\sigma}\mathrm{d}\boldsymbol{w}_t, \\
& \boldsymbol{X}_0 \sim \delta_{x_0}, \quad \boldsymbol{X}_1 \sim \delta_{x_1},
\end{aligned}
\end{equation*}
where $\delta_{x_0}$ and $\delta_{x_1}$ are Dirac delta functions centered at $x_0$ and $x_1$, respectively. 

Assume $\sigma$ is constant and then the corresponding  Fokker–Planck equation in \eqref{eq:sb}~yields
$$
\partial_t p(\boldsymbol{x}, t) = -\nabla_{\boldsymbol{x}} \cdot \left(p(\boldsymbol{x}, t) \boldsymbol{b}(\boldsymbol{x}, t) \right) + \frac{1}{2}\sigma^2 \Delta p(\boldsymbol{x}, t).
$$
From this equation, it can be derived that the ODE
\begin{equation}\label{eq:probflowode}
    \rmd \boldsymbol{X}_t=\underbrace{\left(\boldsymbol{b}\left(\boldsymbol{X}_t, t\right)-\frac{1}{2}\sigma^2 \nabla_{\boldsymbol{x}}\log p \left(\boldsymbol{X}_t, t\right)\right)}_{v(\boldsymbol{X}_t, t)}\rmd t,
\end{equation}
together with the initial distribution generate the same distribution as SDE. The~\eqref{eq:probflowode}  is called the probability flow ODE. Conversely, if~the probability flow ODE $\vxt$ and $\nabla_{\boldsymbol{x}}\log p(\boldsymbol{x}, t)$ (also known as score function) are known, one can recover the SDE drift through $\vxt=\bxt+\frac{1}{2}\sigma^2\nabla_{\boldsymbol{x}}\log \pxt$. So the flow-matching objective is 
$$
\mathcal{L}_{\mathrm{U}[\mathrm{SF}]^2 \mathrm{M}}(\theta)=
\mathbb{E}[\underbrace{\left\|\boldsymbol{v}_\theta(\boldsymbol{x}, t)-\vxt\right\|^2}_{\text {flow matching loss }}+\lambda(t)^2 \underbrace{\left\|\nabla s_\theta(\boldsymbol{x}, t)-\nabla \log p(\boldsymbol{x}, t)\right\|^2}_{\text {score matching loss }}].
$$
However, this loss is intractable; by \eqref{eq:brownianbridge} and the CFM objective, one can transform it into a tractable loss
 \begingroup
\makeatletter\def\f@size{8.5}\check@mathfonts
\def\maketag@@@#1{\hbox{\m@th\normalsize\normalfont#1}}%
$$
\mathcal{L}_{[\mathrm{SF}]^2 \mathrm{M}}(\theta) =\mathbb{E}_{Q^{\prime}} \underbrace{\left\|\boldsymbol{v}_\theta(\boldsymbol{x}, t)-\boldsymbol{v}(\boldsymbol{x}, t \mid (\boldsymbol{x}_0, \boldsymbol{x}_1)\right\|^2}_{\text {conditional flow matching loss }} \\
 +\mathbb{E}_{Q^{\prime}} \lambda(t)^2 \underbrace{\left\|\nabla s_\theta(\boldsymbol{x}, t)-\nabla \log p(\boldsymbol{x}, t \mid (\boldsymbol{x}_0, \boldsymbol{x}_1)\right\|^2}_{\text {conditional score matching loss }},
 $$
 \endgroup
where $Q'=t \sim \mathcal{U}(0,1) \otimes q(\boldsymbol{x}_0, \boldsymbol{x}_1) \otimes  p (\boldsymbol{x}, t |(\boldsymbol{x}_0, \boldsymbol{x}_1))$. Since the conditional path is a Brownian bridge, the~analytic form can be derived, i.e.,~$p(\boldsymbol{x}, t \mid (\boldsymbol{x}_0, \boldsymbol{x}_1) )=\mathcal{N}\left(\boldsymbol{x}; t \boldsymbol{x}_1+(1-t) \boldsymbol{x}_0, \sigma^2 t(1-t)\right)$ and 
$$
\boldsymbol{v}(\boldsymbol{x},t \mid (\boldsymbol{x}_0, \boldsymbol{x}_1))=\frac{1-2 t}{t(1-t)}\left(\boldsymbol{x}-\left(t \boldsymbol{x}_1+(1-t) \boldsymbol{x}_0\right)\right)+\left(\boldsymbol{x}_1-\boldsymbol{x}_0\right),
$$
$$\nabla_{\boldsymbol{x}} \log p(\boldsymbol{x}, t \mid (\boldsymbol{x}_0, \boldsymbol{x}_1) )=\frac{t \boldsymbol{x}_1+(1-t)\boldsymbol{x}_0-\boldsymbol{x}}{\sigma^2 t(1-t)}, ~ t \in [0, 1].$$  
And $q(\boldsymbol{x}_0, \boldsymbol{x}_1)$ can be computed by the entropic optimal~transport. 

\textbf{Unbalanced Wasserstein–Fisher–Rao metric ($g \neq 0,  \sigma = 0$)}
  In this case, the~model can account for the unbalanced 
dynamics, however, it can not account for stochastic dynamics. The~cellular dynamics are also governed by the ODE model 
 by $\rmd \xt=\boldsymbol{b}(\xt, t)\rmd t$. Then, one can use the dynamical unbalanced optimal transport to model these dynamics, also known as Wasserstein–Fisher–Rao metric~\cite{uot1,uot2,liunnormalized}, which can be stated as
 ~\vspace{-6pt}
\begin{equation}\label{eq:WFR}
    \begin{gathered}
\inf _{\left(\pxt, \bxt,\gxt\right)} \int_0^1 \int_{\mathbb{R}^d} \left(\frac{1}{2}\|\bxt\|_2^2 +\alpha |\gxt|_2^2\right)  \pxt \rmd \boldsymbol{x} d t, \\
\mbox{s.t. }\partial_t p+\nabla \cdot\left(\bxt p\right)=g(\boldsymbol{x},t) p,\ p|_{t=0}=\nu_0, p|_{t=1}=\nu_1 .
\end{gathered}
\end{equation}
Here,  $\alpha$  denotes a hyperparameter that controls the weighting. It is also important to note that in this context,  $\nu_0$  and  $\nu_1$  do not necessarily correspond to normalized probability densities; rather, they generally represent mass~densities. 

Recent works such as TrajectoryNet and TIGON utilize \eqref{eq:WFR} to infer unbalanced dynamics from scRNA-seq data~\cite{Tigon,tong2023unblanced}. To~derive a CNF solver for \eqref{eq:WFR}, TIGON~\cite{Tigon} observes that along the characteristic line $\frac{\mathrm{d}\boldsymbol{x}}{\mathrm{d} t}=\bxt$, one has
\[
\int_0^1 \int_{\mathbb{R}^d} f(\boldsymbol{x}, t) p(\boldsymbol{x},t) \mathrm{d}\boldsymbol{x} \mathrm{d} t=\mathbb{E}_{\boldsymbol{x_0} \sim p_0} \int_0^1 f(\boldsymbol{x}, t) e^{\int_0^tg(\boldsymbol{x}, s) \rmd s} \mathrm{d} t
\]
 and $\frac{\mathrm{d}(\ln p)}{\mathrm{d} t}=g-\nabla \cdot \boldsymbol{v}$.  This can make the computation of both energy loss and reconstruction loss in high dimensional space tractable. Therefore, one can parameterize $\bxt$ and $g(\boldsymbol{x}, t)$ using neural networks, respectively, and~train them by minimizing the overall~loss.

 {\bf \textbf{Regularized Unbalanced Optimal Transport ($g \neq 0,  \sigma \neq 0$)}}  In this case, the~model can account for both the unbalanced 
and stochastic dynamics. The~cellular dynamics are governed by the SDE model $\mathrm{d} \xt=\boldsymbol{b}\left(\xt, t\right) \mathrm{d} t+\sigma(t) \boldsymbol{I}\mathrm{d}\boldsymbol{w}_t$.  We can use the regularized unbalanced optimal transport to model it~\cite{DeepRUOT,branRUOT}. It can be viewed as an unbalanced relaxation of the dynamic formulation of the Schrödinger bridge problem. 
     Consider
\begin{equation}\label{Eq:3.3}
\inf _{\left(p, \boldsymbol{b}, g\right)} \int_0^1 \int_{\Rd}\frac{1}{2}\left\|\bxt\right\|_2^2\pxt\rmd \boldsymbol{x}\mathrm{d} t+ \int_0^1 \int_{\Rd} \alpha \Psi\left(g(\boldsymbol{x},t)\right)\pxt    \rmdx\mathrm{d} t,
    \end{equation}
    where {$\Psi: \mathbb{R} \rightarrow [0, +\infty]$ corresponds to the growth penalty function,} the infimum is taken over all pairs $\left(p, \boldsymbol{b}\right)$ such that $p(\cdot, 0)=$ $\nu_0, p(\cdot, 1)=\nu_1, p(\boldsymbol{x},t)$ absolutely continuous, and~\begin{equation}
        \label{eq:undsb3}
\partial_t p=-\nabla_{\boldsymbol{x}} \cdot\left(p \boldsymbol{b}\right)+\frac{1}{2}\nabla_{\boldsymbol{x}}^2: \left(\sigma^2(t)\boldsymbol{I}p\right)+gp 
    \end{equation}
    with vanishing boundary condition: $\displaystyle \lim_{|\boldsymbol{x}| \rightarrow \infty}p(\boldsymbol{x},t)=0$.

One can similarly develop a dynamical OT solver relying on a neural SDE solver, which might be less efficient compared to a neural ODE solver. Recently, DeepRUOT~\cite{DeepRUOT} reformulates the RUOT problem with the Fisher information regularization, equivalently expressed as\vspace{-16pt}
\begin{adjustwidth}{-\extralength}{0cm}
\centering
\begin{equation}
\label{eq:unddsb_fisher}
\inf _{\left(p, \boldsymbol{v}, g\right)} \int_0^1\int_{\Rd}\left[ \frac{1}{2}\left\|\vxt\right\|_2^2+\frac{\sigma^4(t)}{8}\left\|\nabla_{\boldsymbol{x}} \log p\right\|_2^2-\frac{\sigma^2(t)}{2}\left(1+\log p\right)g+ \alpha \Psi\left(g\right) \right]\pxt\rmdx\mathrm{d} t,
\end{equation}
\end{adjustwidth} 
\textls[-25]{where the infimum is taken over all triplets $\left(p, \boldsymbol{v}, g\right)$ such that} $p(\cdot, 0)=$ $\nu_0, p(\cdot, 1)=\nu_1, p(\boldsymbol{x},t)$ absolutely continuous, and~\vspace{-6pt}\begin{equation}
\label{eq:unddsb_fisher_contineq}
    \partial_t p=-\nabla_{\boldsymbol{x}} \cdot\left(p \vxt\right) + g(\boldsymbol{x},t) p 
\end{equation}
with vanishing boundary condition: $\displaystyle \lim_{|\boldsymbol{x}| \rightarrow \infty}p(\boldsymbol{x},t)=0$.
 Here $\vxt$ is a new vector field, representing the probability flow ODE~field.

Thus, the~original SDE
  $\mathrm{d} \xt=\left(\boldsymbol{b}\left(\xt, t\right)\right) \mathrm{d} t+{\sigma}\left(t\right)\mathrm{d}\boldsymbol{w}_t$
 now can be transformed 
  into the probability flow ODE
\vspace{-6pt}
  $$
  \mathrm{d} \xt=\underbrace{\left(\boldsymbol{b}\left(\xt, t\right)-\frac{1}{2} \sigma^2(t)\nabla_{\boldsymbol{x}} \log p(\xt,t)\right)}_{\boldsymbol{v}\left(\xt, t\right)} \rmd t.
  $$
 If the probability flow ODE’s drift $\vxt$, $\sigma(t)$ and the \emph{score function} $\nabla_{\boldsymbol{x}} \log p(\boldsymbol{x},t)$ are specified, then the
the drift term $\boldsymbol{b}(\boldsymbol{x},t)$ of the SDE can be recovered by
$
\boldsymbol{b}(\boldsymbol{x},t)=\vxt+\frac{1}{2}\sigma^2(t) \nabla_{\boldsymbol{x}} \log p(\boldsymbol{x},t).
$
{Therefore, specifying an SDE is equivalent to specifying the probability flow ODE and the score function $\nabla_{\boldsymbol{x}} \log \pxt$}.
One can then use neural networks $\boldsymbol{v}_\theta, ~ g_\theta$ and $s_\theta$ to parameterize $\vxt$, $\gxt$, and~$\frac{1}{2}\sigma^2(t)\log \pxt$, respectively. 

To train DeepRUOT, the~overall loss is composed of three parts, i.e.,~the energy loss, reconstruction loss, and~the Fokker–Planck constraint:\vspace{-6pt}
\begin{equation}\label{eq:total_loss}
\mathcal{L}= \mathcal{L}_{\text{Energy}}+\lambda_r\mathcal{L}_{\text{Recons}}+\lambda_f\mathcal{L}_{\text{FP}}.
\end{equation}
\textls[-20]{The $\mathcal{L}_{\text{Energy}}$ loss aims for the least action of kinetic energy in Equation \eqref{eq:unddsb_fisher}, which can be computed via CNF by adopting the similar approach in TIGON~\cite{Tigon}. The~reconstruction loss $\mathcal{L}_{\text{Recons}}$ aims the  dynamics to match
 data distribution at the later time point (i.e., $p(\cdot, 1)=\nu_1$).} To~achieve the matching in unbalanced settings, DeepRUOT further decomposes it into two parts:
\begin{equation}\label{eq:unddsb_recon}
\mathcal{L}_{\text{Recons}}=\lambda_m\mathcal{L}_{\text{Mass}}+\lambda_d\mathcal{L}_{\text{OT}}
\end{equation}
where $\mathcal{L}_{\text{Mass}}$ aims to align the number of cells and  $\mathcal{L}_{\text{OT}}$ uses normalized weights to perform optimal transport matching. Lastly, $\mathcal{L}_{\text{FP}}$ aims to let the three parameterized neural networks satisfy the Fokker–Planck constraints \eqref{eq:unddsb_fisher_contineq}. DeepRUOT first utilizes a  Gaussian mixture model to estimate the initial distribution, ensuring that it satisfies the initial conditions $p_0$, and~the physics-informed (PINN) loss~\cite{PINN} is defined as\vspace{-6pt}
\begin{equation}\label{eq:fp}
  {\mathcal{L}_\text{FP}}=\left\| \partial_t p_{\theta}+\nabla_{\boldsymbol{x}} \cdot\left(p_{\theta} \boldsymbol{v}_{\theta}\right) - g_{\theta}p_{\theta}\right\|+ \lambda_w\left\| p_{\theta}(\boldsymbol{x},0)-p_0\right\|, p_{\theta}=\exp{\frac{2}{\sigma^2}s_{\theta}}.
\end{equation}

In~\cite{DeepRUOT}, DeepRUOT adopts a two-stage training approach to stabilize the training process. For~the pre-training stage, they use reconstruction loss only to train $\boldsymbol{v}_{\theta}$ and $g_{\theta}$. Then, they fix $\boldsymbol{v}_{\theta}$ and $g_{\theta}$ and employ conditional flow-matching~\cite{cfm_lipman,cfm_tong,sflowmatch} to learn the log density function ($s_\theta(\boldsymbol{x}, t)$). Finally, for~the {training stage}, they use the  $\boldsymbol{v}_{\theta}$, $g_{\theta}$, and~the log density function as the starting point, then obtain the final result by minimizing the total loss \eqref{eq:total_loss}.

\section{Dynamic Modeling of Spatial~Transcriptomics} 
\label{sec:st}
In this section, we review the dynamical modeling approaches for spatial transcriptomics data. We will first present several random walk- or ODE-based methods to model the snapshot spatial data. Next, we focus on the recent progress to dissect the spatiotemporal dynamics underlying datasets with both space and time resolutions. {Figure \ref{fig:spatial} provides an overview of these spatial transcriptomics modeling approaches.}

\begin{figure}[H]

\begin{adjustwidth}{-\extralength}{0cm}
\centering 
        \includegraphics[width=0.95\linewidth]{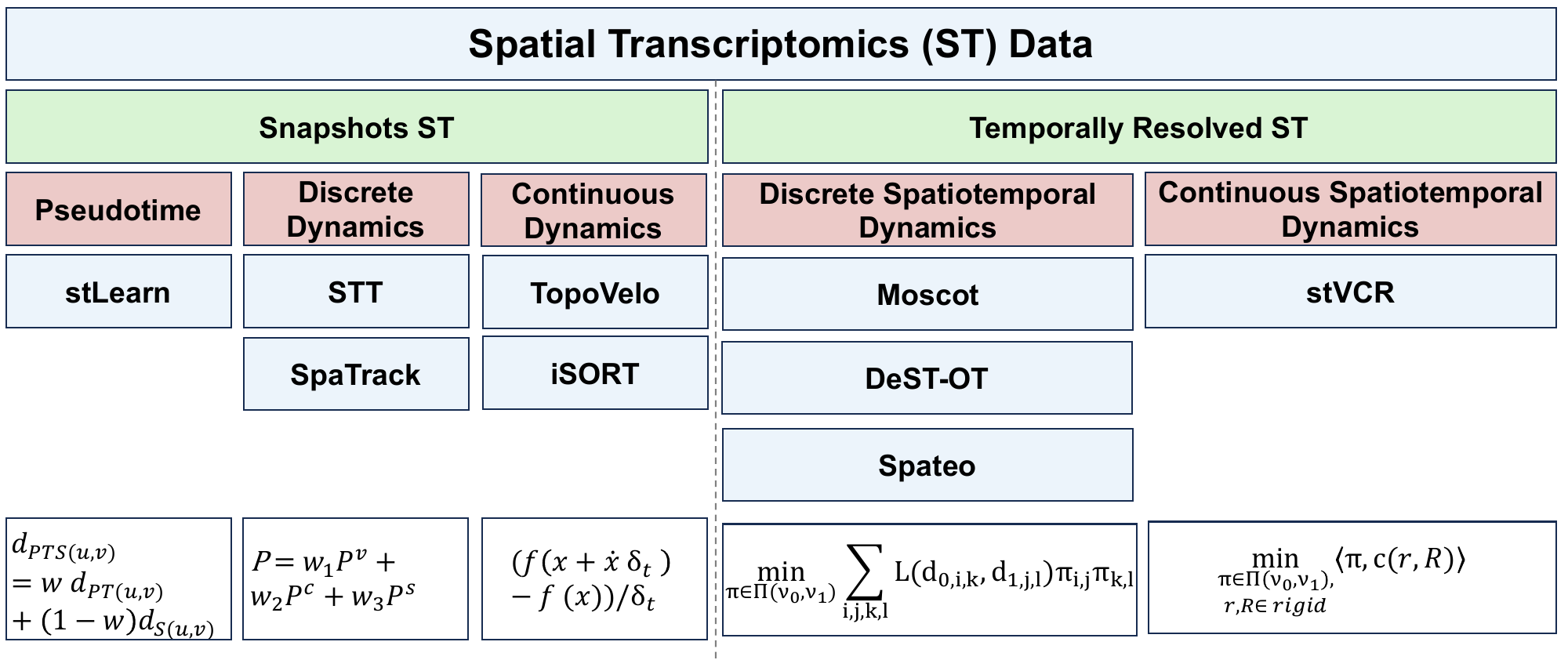}
\end{adjustwidth}
        \caption{\textbf{{Dynamic} 
 modeling of spatial transcriptomics}.}
        \label{fig:spatial}

\end{figure}
\unskip
\subsection{Snapshot Spatial~Transcriptomics}
Below we describe several modeling strategies for single snapshot spatial transcriptomics data, including pseudotime, random walk, and~continuous differential equation \mbox{models,~respectively.}
\subsubsection{Pseudotime~Methods} 
In the context of spatiotemporal trajectory inference, stLearn~\cite{pham2023robust} proposes the PSTS algorithm that combines spatial information and geodesic-based pseudotime information to infer the spatiotemporal developmental trajectory of cells. The~pseudotime distance between two clusters is defined as\vspace{-6pt}
$$
\text{d}_{\text{PT}}(u, v) = \frac{1}{n} \sum_{i=1}^{n} \sum_{j=1}^{n}\left(1 - \frac{p_{u,i} \cdot p_{v,i}}{\|p_{u,i}\| \cdot \|p_{v,j}\|} \right),
$$
where $p_{u,i}$ and $p_{v,i}$ are the PCA vectors of gene expression data points in two clusters. The~spatial distance is defined as the Euclidean distance between the centroids of the two clusters. The~spatiotemporal distance between clusters is the weighted sum of pseudotime distance and spatial distance\vspace{-6pt}
$$
d_{\text{PTS}}(u, v) = \omega d_{\text{PT}}(u, v) + (1 - \omega) d_{S}(u, v).
$$
Each cluster is then treated as a node in a graph, and~the edge weights are determined by $d_{\text{PTS}}$. By~optimizing edge selection using a minimum spanning tree, the~optimal trajectory structure can be~identified. 

{\subsubsection{ Discrete Spatial Dynamics~Modeling}}

STT~\cite{zhou2024spatial} is a random walk-based algorithm to detect multi-stable attractors in spatial transcriptomics. Central to STT is the incorporation of a space coordinate-aware random walk, with~the transition probability matrix having the form $P = w_1P_v+w_2P_c+(1-w_1-w_2)P_s$, where $P_v$ is induced by an attractor-specific RNA velocity (named spatial transition tensor), $P_c$ is induced by gene expression similarity (i.e., diffusion in the gene space), and~$P_s$ is induced by space coordinates (i.e., diffusion in the physical space). By~iteratively \linebreak (1) decomposing $P$ to identify attractors and assign attractor membership to each individual cell and (2) improving attractor-specific RNA velocity estimation, STT is able to identify transitional cells in the snapshot spatial transcriptomics data and plot the local streamlines within~attractors.

SpaTrack~\cite{shen2025inferring} is a spatial transcriptomics analysis tool based on optimal transport theory, which reconstructs cell differentiation trajectories by integrating gene expression profiles and spatial coordinates of cells. When processing Snapshot data, SpaTrack defines the transition cost matrix between cells by weighting gene expression distance and spatial distance as follows:
$
C_{ij} = \alpha_{1} \|\boldsymbol{g}_{i} - \boldsymbol{g}_{j}\|^{2} + \alpha_{2} \|\boldsymbol{z}_{i} - \boldsymbol{z}_{j}\|^{2}
$
where $\boldsymbol{g}_{i}$ represents the gene expression of cell $i$, $\boldsymbol{x}_{i}$ denotes the spatial coordinates of cell $i$, and~$\alpha_{1},\alpha_{2}$ are weighting coefficients. The~transition probability matrix between cells is obtained by solving the following entropy-regularized optimal transport (OT) problem
\begin{equation*}
\boldsymbol{P} = \arg \min_{\boldsymbol{P}} \sum_{ij} C_{ij}P_{ij} + \epsilon H(\boldsymbol{P}) \quad \mathrm{s.t.} \sum_{i}P_{ij} = 1, \sum_{j}P_{ij} = 1
\end{equation*}
where $H(\boldsymbol{P})$ denotes the entropy regularization term and $\epsilon$ is the regularization coefficient. SpaTrack identifies trajectory starting points using single-cell entropy. Let the identified starting points be cells $1,2,\cdots,s$. The~probability of transitioning from starting cells to cell $i$ can be calculated as $\gamma_{i} = \sum_{j=1}^{s} P_{ji}$. By~sorting cells in ascending order based on their $\gamma_{i}$ values, the~position of each cell in the differentiation trajectory can be~determined.

\vspace{1em}
{\subsubsection{Continuous Spatial Dynamics~Modeling}}
Several methods aim to extend the continuous RNA velocity model of scRNA-seq data toward snapshot spatial transcriptomics. One recent method, iSORT~~\cite{Jifan_isort} uses transfer learning to obtain a mapping of gene expression to spatial location $\boldsymbol{z} = f(\boldsymbol{x})$ and proposes the concept of spatial RNA velocity, which utilizes the velocity of gene expression and the mapping $\boldsymbol{z} = f(\boldsymbol{x})$ to obtain spatial RNA velocity, formally
\begin{equation*}
    \frac{\rmd \boldsymbol{z}}{\rmd t} = \nabla f \cdot\frac{\rmd \boldsymbol{x}}{\rmd t}.
 \end{equation*}

In addition, Topovelo~~\cite{Topovelo} uses a graph neural network to infer RNA velocity for spatial transcriptomics data and suggests that a decoder could be trained to further infer continuous spatial~velocities.

\vspace{1em}
\subsection{Temporally Resolved Spatial~Transcriptomics}
The availability of time-series ST data opens new avenues to explore cellular migration within physical space~\cite{peng2024stvcr,qiu2022spateo,Topovelo}. Nevertheless, the~inherently destructive nature of sequencing limits ST data to static snapshots rather than continuous trajectories. Particularly, when sequencing is performed at various time points during embryonic development, the~resulting time-series ST data are often derived from distinct biological samples, leading to multiple unpaired snapshots~\cite{chen2022spatiotemporal,wei2022single,wang2024single}. In~addition, due to possible rotation, translation, and~stretching of different slices, the~spatial coordinates of different samples are not in the same coordinate system~\cite{PASTE,qiu2022spateo,peng2024stvcr}. Therefore, reconstructing trajectories of cell state transition, proliferation, and~migration for time-series ST data is a challenging~task.

To overcome these challenges, many methods have been developed in recent years. Similar to modeling temporal single-cell data, these methods can be divided into two categories: those that model dynamics on discrete cell states~\cite{moscot,DestOT,qiu2022spateo} and those that model dynamics in continuous spaces~\cite{peng2024stvcr}.

{\subsubsection{ Discrete Spatiotemporal  Dynamics~Modeling}}
Among the methods that model spatiotemporal dynamics on discrete cell states, recent work includes Moscot~\cite{moscot}, DeST-OT~\cite{DestOT}, and~Spateo~\cite{qiu2022spateo}. These approaches employ fused Gromov–Wasserstein optimal transport~\cite{titouan2019optimal,chowdhury2019gromov} as the main tool, which was first used by PASTE~\cite{PASTE} to align adjacent 2D slices to reconstruct the 3D structure of the~tissue.

{\bf \textbf{Fused Gromov–Wasserstein Optimal Transport}}
We consider two adjacent unpaired slices $(\boldsymbol{X},\boldsymbol{Z})$ and $(\boldsymbol{X}^\prime,\boldsymbol{Z}^\prime)$ with spots (or cell) numbers $N$ and $M$, where $\boldsymbol{X} \in \mathbb{R}^{N \times G}$, $\boldsymbol{X}^\prime \in \mathbb{R}^{M \times G}$ are the gene expression of the two slices, and~$\boldsymbol{Z}\in \mathbb{R}^{N \times 2}$ and $\boldsymbol{Z}^\prime\in \mathbb{R}^{M \times 2}$ are the spatial coordinates. In~addition, the~spatial coordinates of each slice can be converted into a distance matrix $\boldsymbol{D} \in \mathbb{R}^{N \times N}_{+}$, where $d_{ij} = \|\boldsymbol{z}_i - \boldsymbol{z}_j\|_2$. The~fused Gromov–Wasserstein optimal transport problem reads as follows:
\begin{equation}
    \begin{aligned}\label{eq:static_fgwot}
     \min_{\boldsymbol{\pi} \in \boldsymbol{\Pi}(\nu_0, \nu_1)} (1-\alpha)\sum_{i, j}c_{i, j}\pi_{i, j} + \alpha \sum_{i, j, k, l}(d_{i,k} - d_{j,l}^\prime)^2\pi_{i, j}\pi_{k, l},
    \end{aligned}
\end{equation}
where $\pi, \boldsymbol{\Pi}(\nu_0, \nu_1)$ and $c$ have the same meaning as before, and~$\alpha$ is a hyperparameter that weighs the importance of gene expression and spatial location. The~fused Gromov–Wasserstein optimal transport (FGW-OT) problems can also be solved by calling the {POT} (version 0.9.5) 
 package~\cite{flamary2021pot}.

{\bf \textbf{Generalized Weighted Procrustes Problem}}
 When the mapping $\boldsymbol{\pi}$ is found, in~order to unify the spatial coordinates of adjacent slices into the same coordinate system by a rigid body transformation (rotation and translation), we need to solve a generalized weighted Procrustes problem. Formally,\vspace{-6pt}
\begin{equation} \label{eq:gwpp}
    \hat{\boldsymbol{R}}, \hat{\boldsymbol{r}} = \mathop{\arg\min}_{\substack{\boldsymbol{R} \in \mathbb{R}^{2 \times 2}, \boldsymbol{r} \in \mathbb{R}^{2} \\ \boldsymbol{R}^T \boldsymbol{R}=\boldsymbol{I}, \det \boldsymbol{R}=1}} \sum_{i, j} \pi_{i j}\left\|\boldsymbol{z}_i-(\boldsymbol{R} \boldsymbol{z}^{\prime}_j +\boldsymbol{r})\right\|^2_2,
\end{equation}
where $\boldsymbol{R}$ is the rotation matrix and $\boldsymbol{r}$ is the translation~vector.

{\bf \textbf{Applications in ST Data}}
Moscot~\cite{moscot} extends FGW-OT to model slices of adjacent time points by adding the entropy regularization mentioned in Equation~\eqref{eq:ethropicot} and the unbalanced settings mentioned in Equations~\eqref{eq:static_ubot} and \eqref{eq:static_ubot_v0}. Formally,
\begin{equation*}
    \begin{aligned}\label{eq:static_fgwot2}
     \min_{\boldsymbol{\pi} \in \mathbb{R}^{N \times M}_{+}} & (1-\alpha)\sum_{i, j}c_{i, j}\pi_{i, j} + \alpha \sum_{i, j, k, l}(d_{i,k} - d_{j,l}^\prime)^2\pi_{i, j}\pi_{k, l}\\
     & + \tau_1 \operatorname{KL}(\boldsymbol{\pi} \boldsymbol{1}_M||\nu_0)+\tau_2 \operatorname{KL}(\boldsymbol{\pi}^T \boldsymbol{1}_N||\nu_1) - \varepsilon H(\boldsymbol{\pi}),
    \end{aligned}
\end{equation*}
where $\tau_1,\tau_2,\varepsilon$ and $H(\boldsymbol{\pi})$ have the same meaning as before and $\nu_0$ is calculated from a pre-selected gene set according to Equation~\eqref{eq:static_ubot_v0}. Note that when we talk about ST data at different time points, the~spatial coordinates can not only be 2D but can also be reconstructed in 3D. We use $d_{\text{spa}}$ to refer to the dimension of the spatial~coordinates.

DeST-OT~\cite{DestOT} designs methods that enable simultaneous inference of cell growth rate and mapping from data. The~DeST-OT optimization problem with the semi-relaxed constraints and entropic regularization is
\begin{equation*}
\begin{array}{ll}
\min & \mathcal{E}_{\text {DeST-OT }}(\boldsymbol{\pi})+\tau_1 \operatorname{KL}(\boldsymbol{\pi} \boldsymbol{1}_M||\nu_0) - \varepsilon H(\boldsymbol{\pi}) \\
\text { s.t. } & \boldsymbol{\pi}^T \boldsymbol{1}_{N}=\nu_1, \quad \boldsymbol{\pi} \in \mathbb{R}^{N \times M}_{+},
\end{array}
\end{equation*}
where $\mathcal{E}_{\text {DeST-OT }}(\boldsymbol{\pi})$ includes the Wasserstein OT term and another term $\mathcal{E}^{\boldsymbol{M}}(\boldsymbol{\pi})$ related to growth and GW-OT, that is,\vspace{-6pt}
\begin{equation*}
    \mathcal{E}_{\text {DeST-OT }}:=(1-\alpha) \sum_{i, j^\prime} c_{i, j^\prime}\pi_{i, j^\prime}+\alpha \mathcal{E}^{\boldsymbol{M}}(\boldsymbol{\pi}).
\end{equation*}
$\mathcal{E}^{\boldsymbol{M}}(\boldsymbol{\pi})$ is defined as 
\vspace{-12pt}
\begin{equation*}
\begin{aligned}
\mathcal{E}^{\boldsymbol{M}}(\boldsymbol{\pi}):=\frac{1}{2}&\left(\sum_{i, j^{\prime}, k^{\prime}} \pi_{i j^{\prime}} \pi_{i k^{\prime}} \boldsymbol{M}_{j^{\prime} k^{\prime}}^{\prime 2}+\sum_{i, j, k^{\prime}} \pi_{i k^{\prime}} \pi_{j k^{\prime}} \boldsymbol{M}_{i j}^{2} \right.\\
&\left.+\sum_{i, j^{\prime}, k, l^{\prime}}\left(\boldsymbol{M}_{i k}-\boldsymbol{M}_{j^{\prime} l^{\prime}}^{\prime}\right)^2 \pi_{i j^{\prime}} \pi_{k l^{\prime}}\right),
\end{aligned}
\end{equation*}
where $\boldsymbol{M} = \boldsymbol{D}^{\text{spa}}\odot\boldsymbol{D}^{\text{exp}}$ measures the distance between two cells in the same slice, and~$\boldsymbol{D}^{\text{spa}}$ and $\boldsymbol{D}^{\text{exp}}$ are distance matrices constructed on each slice according to spatial coordinates $\boldsymbol{S}$ and gene expression $\boldsymbol{X}$, respectively. That is, $d^{\text{spa}}_{i,j} = \|\boldsymbol{z}_i-\boldsymbol{z}_j\|_2$ and $d^{\text{exp}}_{i,j} = \|\boldsymbol{x}_i-\boldsymbol{x}_j\|_2$. In~addition, the~first and second terms in $\mathcal{E}^{\boldsymbol{M}}(\boldsymbol{\pi})$ promote the proximity of different descendants of a cell at the previous moment and the proximity of different ancestors of a cell at the later moment, respectively, and~the third term is the usual GW term that only replaces the spatial distance matrix $\boldsymbol{D}$ with the $\boldsymbol{M}$ distance matrix. In~DeST-OT, the~authors define the growth vector $\boldsymbol{\xi}=\boldsymbol{\pi} \boldsymbol{1}_{M}-\nu_0$ and the growth rate $\boldsymbol{g}=\log \left(1+N \boldsymbol{\xi}\right) /\left(t_1-t_0\right)$. 

Spateo~\cite{qiu2022spateo} uses maps $\pi$ obtained by other OT-based methods to unify spatial coordinates at different times into the reference coordinate system by solving the generalized weighted Procrustes problem in Equation~\eqref{eq:gwpp} (possibly in 3D). Next, Spateo selects the spatial coordinates of the cell with the most weight at the late time point mapped from each cell at the early time point as its future state, formally\vspace{-6pt}
$$
\boldsymbol{z}_{i, \text{future}} = \boldsymbol{z}^{\prime}_{\arg\max\pi_{i,:}},
$$
where $\pi_{i,:}$ refers to the $i$ row of $\boldsymbol{\pi}$, that is, the~weight of the $i$ cell mapped from the early time point to each cell at the late time point. When the future spatial coordinates of each cell are determined, we can define the spatial velocity of each cell
$$
\boldsymbol{v}^{\text{spa}}_{i} = \boldsymbol{z}_{i, \text{future}} - \boldsymbol{z}_i.
$$
Finally, Spateo recovers a continuous spatial velocity field 
$
\frac{\rmd \boldsymbol{z}}{\rmd t} = f(\boldsymbol{z}),
$
from the spatial velocity of each cell, allowing for a series of differential geometry analyses, including divergence, acceleration, curvature, and~torsion.

\vspace{12pt}
\subsubsection{Spatiotemporal  Dynamics~Modeling}

The majority of current approaches model spatial coordinates based on Gromov–Wasserstein OT, which has no dynamic form. Recently, stVCR~\cite{peng2024stvcr} proposes to model spatial coordinates using rigid-body transformation invariant OT, as~well as using the widely used Wasserstein OT for modeling gene expression and unbalanced OT for modeling cellular proliferation. Next, stVCR integrates all modules into dynamic forms, making it possible to reconstruct dynamic continuous trajectories of cell differentiation, migration, and~proliferation~simultaneously.

\emph{Rigid body transformation invariant optimal transport}
The method in~\cite{cohen1999earth} considers the optimal transport problem invariant to a given set of manipulations $\mathcal{G}$. It simultaneously searches for the optimal mapping $\pi$ and the optimal transformation $g$ through the optimization problem:\vspace{-6pt}
\begin{equation}\label{eq:rtiot}
    (\boldsymbol{\pi}^{\star}, g^{\star}) = \mathop{\arg\min}_{\boldsymbol{\pi} \in \boldsymbol{\Pi}(\nu_0, \nu_1), g \in \mathcal{G}} \langle \boldsymbol{c}(g), \boldsymbol{\pi}\rangle \stackrel{\text { def. }}{=}  \sum_{i,j} \pi_{i,j} d\big(\boldsymbol{z}_i, g(\boldsymbol{Z}^\prime_j)\big). 
\end{equation}
Solving problem~\eqref{eq:rtiot} directly is difficult, and~it can be solved iteratively by
\vspace{-6pt}
\begin{align}
    \boldsymbol{\pi}^{(n)}&=\mathop{\arg\min}_{\boldsymbol{\pi} \in \boldsymbol{\Pi}(\nu_0, \nu_1)} \sum_{i,j} \pi_{i j} d\big(\boldsymbol{z}_i, g^{(n)}(\boldsymbol{Z}^\prime_j)\big), \label{iOT sub1} \\
    g^{(n+1)}&=\mathop{\arg\min}_{g \in \mathcal{G}} \sum_{i,j} \pi_{i j}^{(n)} d\big(\boldsymbol{z}_i, g(\boldsymbol{Z}^\prime_j)\big) \label{iOT sub2}.
\end{align}
The subproblem~\eqref{iOT sub1} is to solve a static OT. In~addition, when we choose the set $\mathcal{G}$ as the set of rigid body transformations, we call this problem rigid body transformation invariant optimal transport. At~this point, subproblem~\eqref{iOT sub2} is the generalized weighted Procrustes problem mentioned~before. 

Consider the ST data $(\boldsymbol{Z}^{(0:K)},\boldsymbol{X}^{(0:K)})$ at $t_0, t_1 \dots t_K$ totaling $K$ time points, and~the number of cells in each observation is $n_0, n_1 \dots n_K$. stVCR uses the spatial coordinate system of the data at $t_0$ as a reference, and~searches for the optimal dynamics and the optimal rigid-body transformation $(\boldsymbol{r}_{1:k}, R_{1:k})$ by interpolating the empirical probability distributions of the data after the rigid-body transformation $\hat{p}^{k}$ and the number of cells $n_k$ using a transport-with-growth partial differential equation (PDE)
\begin{equation}\label{eq:twgPDE}
    \partial_t p_t(\boldsymbol{z}, \boldsymbol{x})+\nabla \cdot\Big(\big(\boldsymbol{v}_t(\boldsymbol{z}, \boldsymbol{x}), \boldsymbol{b}_t(\boldsymbol{z}, \boldsymbol{x})\big) p_t(\boldsymbol{z}, \boldsymbol{x})\Big)=g_t(\boldsymbol{z}, \boldsymbol{x}) p_t(\boldsymbol{z}, \boldsymbol{x}),
\end{equation}
where $\boldsymbol{v}_t(\boldsymbol{z}, \boldsymbol{x})$ is cell spatial migration velocity, $\boldsymbol{b}_t(\boldsymbol{z}, \boldsymbol{x})$ is RNA velocity and $g_t(\boldsymbol{z}, \boldsymbol{x})$ is cell growth rate. Thus, the~feasible state space $\mathcal{S}$ for the arguments under constraints is
\begin{equation}
    \begin{aligned}
        \mathcal{S}(&\boldsymbol{Z}^{(0:K)}, \boldsymbol{X}^{(0:K)})  := \big\{(p_t, \boldsymbol{v}_t, \boldsymbol{b}_t,  g_t; \boldsymbol{R}_{1:K}, \boldsymbol{r}_{1:K} ) \big|\ \partial_t p_t+\nabla \cdot((\boldsymbol{v}_t, \boldsymbol{b}_t) p_t)=g_t p_t, \\
        & p_{t_0}=p^{(0)}, \|p_{t_k}\|_1 = n_k/n_0, \bar{p}_{t_k}={p}^{(k)},  \boldsymbol{R}^{T}_k \boldsymbol{R}_k=\boldsymbol{I}, \det \boldsymbol{R}_k=1,\ k=1,2,\ldots, K  \big\}, 
    \end{aligned}\label{whole ot cons}
\end{equation}
where $\|p_t\|_1:=\int p_t \mathrm{d}\boldsymbol{z}\mathrm{d}\boldsymbol{x}$ is the total mass of $p_t$ and $\bar{p}_t := p_t/\|p_t\|_1$. stVCR finds optimal dynamics $(p_t, \boldsymbol{v}_t, \boldsymbol{b}_t,  g_t)$ and optimal rigid-body transformations $(\boldsymbol{R}_{1:K}, \boldsymbol{r}_{1:K})$ by minimizing the Wasserstein–Fisher–Rao (WFR) distance
\begin{equation}
    \int_{t_0}^{t_K} \int_{\mathbb{R}^{G+d_{\text{spa}}}} \left(\left\|\boldsymbol{v}_t\right\|^2 + \alpha_{\text{Exp}} \left\|\boldsymbol{b}_t\right\|^2 + \alpha_{\text{Gro}} g^2_t\right)  p_t(\boldsymbol{z},\boldsymbol{x}) \mathrm{d}\boldsymbol{z}\mathrm{d}\boldsymbol{x}\mathrm{d} t
    \label{eq:wfr_spa}
\end{equation}
for $(p_t, \boldsymbol{v}_t, \boldsymbol{b}_t,  g_t; \boldsymbol{R}_{1:K}, \boldsymbol{r}_{1:K}) \in \mathcal{S}(\boldsymbol{Z}^{(0:K)}, \boldsymbol{X}^{(0:K)})$. According to the direct derivation of the solution of the Feynman–Kac type PDE~\eqref{eq:twgPDE} by characteristics, Equation~\eqref{eq:wfr_spa} has a dimensionally independent form\vspace{-6pt}
\begin{equation}\label{whole ot2}
    \begin{aligned}
        \mathcal{L}_{\text{Dyn}} =\mathbb{E}_{(\boldsymbol{z}^{(t_0)}, \boldsymbol{x}^{(t_0)})\sim p^{(0)}}\int_{t_0}^{t_K} \Big(&\|\boldsymbol{v}_t(\boldsymbol{z}^{(t)}, \boldsymbol{x}^{(t)})\|^2 + \alpha_{\text{Exp}} \|\boldsymbol{b}_t(\boldsymbol{z}^{(t)}, \boldsymbol{x}^{(t)})\|^2 \\
        &+ \alpha_{\text{Gro}} \|g_t(\boldsymbol{z}^{(t)}, \boldsymbol{x}^{(t)})\|^2\Big) w_t[\boldsymbol{z},\boldsymbol{x}]\mathrm{d} t,
    \end{aligned}
\end{equation}
where $\boldsymbol{z}^{(t)}, \boldsymbol{x}^{(t)}, w_t[\boldsymbol{z},\boldsymbol{x}]$ satisfies the characteristic ordinary differential equations (ODEs)
\begin{equation}\label{eq:Char-ODE}
    \begin{aligned}
        \frac{d\boldsymbol{z}^{(t)}}{dt}=\boldsymbol{v}_t(\boldsymbol{z}^{(t)},\boldsymbol{x}^{(t)}),\ \frac{d\boldsymbol{x}^{(t)}}{dt}&=\boldsymbol{b}_t(\boldsymbol{z}^{(t)},\boldsymbol{x}^{(t)}),\  
        \frac{d \ln w_t}{d t}= g_t(\boldsymbol{z}^{(t)},\boldsymbol{x}^{(t)}), \\
        (\boldsymbol{z}^{(t)},\boldsymbol{x}^{(t)}, w_t)|_{t=t_0}&=(\boldsymbol{z}^{(t_0)},\boldsymbol{x}^{(t_0)}, 1).
    \end{aligned}
\end{equation}
stVCR implemented the constraints $\|p_{t_k}\|_1 = n_k/n_0$ and  $\bar{p}_{t_k}=\hat{p}^{(k)}$ in Equation~\eqref{whole ot cons} as soft penalties by performing distribution matching\vspace{-6pt}
\begin{equation}\label{distance match}
    \mathcal{L}_{\text{Mch}}=\sum_{k=1}^K \big(W_2(\bar{p}_{t_k}, \hat{p}^{(k)})\big)^2+ \kappa_{\text{Gro}} \sum_{k=1}^K \frac{|\sum_{j=1}^{n_0} w_{t_k,j} - n_k|}{n_k},
\end{equation}
where the second term promotes a reduction in the relative error of the total mass, and~the first term is the 2-Wasserstein distance between the normalized distribution corresponding to the dynamics $\bar{p}_{t_k}$ and the probability distribution of the observed data after rigid body transformation $\hat{p}^{(k)}$, where the cost function is defined as
$$
c^{(k)}_{ij}  = \kappa_{\text{Exp}} \|\boldsymbol{x}^{(t_k)}_i -\hat{\boldsymbol{x}}^{(k)}_j\|_2^2 + (1-\kappa_{\text{Exp}}) \|\boldsymbol{z}^{(t_k)}_i - \hat{\boldsymbol{z}}^{(k)}_j\|_2^2,\quad i=1:n_0, j=1:n_k,
$$ 
where $\kappa_{\text{Exp}}$ weighs the importance of gene expression and spatial coordinates in distribution matching. In~addition, for~annotated data, stVCR achieves modeling of known type transitions by modifying $\mathcal{L}_{\text{Mch}}$, and~spatial structure preservation for specified organs or tissues by adding an optional objective function $\mathcal{L}_{\text{SSP}}^{\text{opt}}$. 

\textls[-25]{In summary, the~loss function of stVCR contains two required items and one optional item}\vspace{-6pt}
\begin{equation}\label{distance match2}
    \mathcal{L} = \mathcal{L}_{\text{Dyn}} + \lambda_{\text{Mch}}\mathcal{L}_{\text{Mch}} + \lambda_{\text{SSP}}\mathcal{L}_{\text{SSP}}^{\text{opt}}.
\end{equation}
In practice, stVCR parameterizes the dynamics $\boldsymbol{v}_t(\boldsymbol{z}^{(t)},\boldsymbol{x}^{(t)})$, $\boldsymbol{b}_t(\boldsymbol{z}^{(t)},\boldsymbol{x}^{(t)})$ and $g_t(\boldsymbol{z}^{(t)},\boldsymbol{x}^{(t)})$ into neural networks as well as parameterizes the rotation matrix $\boldsymbol{R}_{1:K}$ into rotation angles $\alpha_{1:K}$ (or Euler angles $(\alpha_{1:K}, \beta_{1:K}, \gamma_{1:K})$ in 3D) and solves iteratively using back-propagation~algorithm.

\section{Extensions, Challenges, and~Future~Directions}\label{sec:extensions}
Recent advancements in single-cell transcriptomics, spatial transcriptomics, and~computational modeling have significantly improved our ability to reconstruct cellular dynamics. However, several outstanding challenges remain, particularly in integrating different discrete and continuous models, handling the complexity of single-cell dynamics, and~ensuring the biological interpretability of inferred dynamical systems. This section discusses key areas for further development, focusing on new methodologies that combine discrete and continuous modeling approaches, the~construction of comprehensive dynamical frameworks, and~the broader applications for modeling cellular fate~decisions.

\subsection{Bridging Discrete and Continuous Dynamics~Modeling}\label{sec:bridge}

One interesting topic to explore is building connections between discrete dynamic models (e.g., Markov chain) with continuous differential equations when dealing with scRNA-seq data. In~CellRank~\cite{lange2022cellrank,weiler2024cellrank}, the~output from continuous models could help to refine the random walk on the data point cloud by introducing various kernels.  Let the spliced RNA counts of cells $i$ and $j$ be $\boldsymbol{s}_i$ and $\boldsymbol{s}_j$, respectively.  The~\textbf{Gaussian Kernel} is defined as
\begin{equation*}
    d_g(\boldsymbol{s}_i, \boldsymbol{s}_j) = \exp\left(-\frac{\|\boldsymbol{s}_i - \boldsymbol{s}_j\|^2}{\sigma^2}\right),
\end{equation*}
Let the position vector from cell $i$ to cell $j$ be $\boldsymbol{\delta}_{ij} = \boldsymbol{s}_j - \boldsymbol{s}_i$ and $v_i$ denotes the estimated RNA velocity of cell $i$. Then, the~three velocity kernels can be introduced~as
\begin{itemize}
    \item Cosine Kernel: $v_{\cos}(\boldsymbol{s}_i, \boldsymbol{s}_j) = g\left(\cos(\boldsymbol{\delta}_{ij}, \boldsymbol{v}_i)\right)$,
    \item Correlation Kernel: $v_{\text{corr}}(\boldsymbol{s}_i, \boldsymbol{s}_j) = g\left(\mathrm{corr}(\boldsymbol{\delta}_{ij}, \boldsymbol{v}_i)\right)$,
    \item Inner Product Kernel: $v_{\text{ip}}(\boldsymbol{s}_i, \boldsymbol{s}_j) = g\left(\boldsymbol{\delta}_{ij}^T \boldsymbol{v}_i\right)$.
\end{itemize}

\noindent {Here,} 
 $g$ is a bounded, positive, monotonic increasing function such as an exponential function. In~CellRank~\cite{lange2022cellrank}, the~actual transition kernel can combine these two parts either by weighted summation or multiplication. For~example, the~kernel used in the original RNA velocity method is\vspace{-6pt}
\begin{equation*}
    \mathrm{Ker}(\boldsymbol{s}_i, \boldsymbol{s}_j) = \lambda v_{\cos}(\boldsymbol{s}_i, \boldsymbol{s}_j) + (1 - \lambda) d_g(\boldsymbol{s}_i, \boldsymbol{s}_j),
\end{equation*}
where $\lambda$ is a weighting coefficient. The~Markov chain transition matrix is constructed as\vspace{-6pt}

\begin{equation*}
    p_{ij} = \frac{\mathrm{Ker}(\boldsymbol{s}_i, \boldsymbol{s}_j)}{\sum_j \mathrm{Ker}(\boldsymbol{s}_i, \boldsymbol{s}_j)}.
\end{equation*}
CellRank2~\cite{weiler2024cellrank} provides more flexible options for the transition kernel and incorporates prior knowledge. For~example, if~pseudotime is known in advance, it can be used to adjust the transition kernel
\begin{equation*}
    \mathrm{Ker}_{\text{adj}}(\boldsymbol{s}_i, \boldsymbol{s}_j) = \mathrm{Ker}(\boldsymbol{s}_i, \boldsymbol{s}_j) f(\Delta t_{ij}),
\end{equation*}
where
\begin{equation*}
    f(\Delta t) = 
    \begin{cases} 
        \frac{2}{\sqrt{1 + \exp(b \Delta t)}} & \Delta t < 0, \\
        1 & \Delta t \geq 0.
    \end{cases}
\end{equation*}
Additionally, a~unified transition kernel can be constructed for multi-time point data. For~data at different time points, a~transport map such as optimal transport (OT) can be used to define $\pi_{t_j, t_{j+1}}$. By~placing the same time point data on the diagonal of a global transition matrix and the transport map between different time points on the off-diagonal, a~global transition matrix $T$ can be obtained, enabling the construction of a Markov chain across different time~points.

\textls[-25]{Another direction is to analyze the theoretical convergence of discrete dynamics as the number of data points tends to infinity. A~well-known example is the study of the continuum limit of diffusion map random walk~\cite{coifman2008diffusion}, stating that the random walk induced by the Gaussian kernel would converge to the dynamics of the Fokker–Planck equation. When considering growth, the~directed random walk defined by PBA would converge to \eqref{eq:pde}.
Interestingly,~{ref.}~\cite{zhou2021dissecting} also proves that the coarse-grained transition probabilities yield the continuum limit of transition rate among attractors \eqref{eq:transitionrate}, therefore validating the rationale of~MuTrans.}

Once such a theoretical connection is built, new theoretical insights could be drawn toward the algorithm design. For~instance,~{ref.}~\cite{li2020mathematics} systematically investigates the continuous limit of RNA-velocity-induced random walk kernels. For~example, if~the transition kernel is
$
    \mathrm{Ker}(\boldsymbol{s}_i, \boldsymbol{s}_j) = d_g(\boldsymbol{s}_i, \boldsymbol{s}_j) \cdot v_{\cos}(\boldsymbol{s}_i, \boldsymbol{s}_j),
$
the corresponding ODE  yields the desired streamlined equation that correctly reveals the vector field directionality
$
    \frac{\mathrm{d}\boldsymbol{x}}{\mathrm{d}t} = \frac{\boldsymbol{v}}{\|\boldsymbol{v}\|}.
$
Meanwhile, for~$\mathrm{Ker}(\boldsymbol{s}_i, \boldsymbol{s}_j) = d_g(\boldsymbol{s}_i, \boldsymbol{s}_j) \cdot v_{\text{corr}}(\boldsymbol{s}_i, \boldsymbol{s}_j)$, the~corresponding ODE is
$
    \frac{\mathrm{d}\boldsymbol{x}}{\mathrm{d}t} = \frac{\mathcal{P}_{\boldsymbol{1}} \boldsymbol{v}}{\|\mathcal{P}_{\boldsymbol{1}} \boldsymbol{v}\|},
$
where $\mathcal{P}$ is the projection operator defined as
$
    \mathcal{P}_n \boldsymbol{x} = (I - \boldsymbol{\hat{n}} \otimes \boldsymbol{\hat{n}}) \cdot \boldsymbol{x}, \quad \boldsymbol{\hat{n}} = \frac{\boldsymbol{n}}{\|\boldsymbol{n}\|}, \quad \boldsymbol{1} = (1, 1, \cdots, 1)^T.
$
which indicates that the correlation kernel might alter both the direction and magnitude of RNA velocity in the continuous~limit.

Some approaches also use discrete graphs to represent the geometric structure of data~\cite{hetzel2021graph}. Graphdynamo~\cite{graph_dynamo} and Graphvelo~\cite{chen2024graphvelo} propose a method that leverages geometric structure to correct RNA velocity. They assume that the cell data points $\boldsymbol{x}_i$ lie on a low-dimensional manifold embedded in a high-dimensional space (in classical mechanics, this is known as the “configuration manifold”) and that each cell’s RNA velocity vector lies in the tangent space $T_x \mathcal{M}$ at the point $\boldsymbol{x}_i$ on the manifold. Let $\boldsymbol{\delta}_{ij}$ denote the displacement vector from cell $i$ to its neighboring cell $j$. With~a sufficient number of such $\boldsymbol{\delta}_{ij}$, one can construct a non-orthogonal normalized basis for $T_x \mathcal{M}$. Thus, the~RNA velocity vector in $T_x \mathcal{M}$ can be expressed as
$
\boldsymbol{v}_{\parallel}(\boldsymbol{x}_i) = \sum_{j \in \mathcal{N}_i} \phi_{ij} \boldsymbol{\delta}_{ij}
$ The coefficients of the linear combination, $\boldsymbol{\phi}_i = \{\phi_{ij} \mid j \in \mathcal{N}_i\}$, are determined by minimizing the following loss:
$$
\mathcal{L}(\phi_i) = \|\boldsymbol{v}_i - \boldsymbol{v}_{\parallel}(\boldsymbol{x}_i)\|^2 - b \cos(\boldsymbol{\phi}_i, \boldsymbol{\phi}_i^{\text{corr}}) + \lambda \|\boldsymbol{\phi}_i\|^2
$$
Here, $\phi_i^{\text{corr}}$ denotes the transition probabilities provided by the Cosine Kernel, and~the last term is a regularization term. Thus, $v_\parallel$ serves as a geometry-aware correlation of $v_i$, ensuring greater coherence with the underlying manifold~structure.

Furthermore, the~population dynamics in the feature space can be transferred to the dynamics on the graph. The~unbalanced Fokker–Planck Equation \eqref{eq:pde} could be generalized to a graph, such that the mass evolution at node $i$ is given by the following equation:
\begingroup
\makeatletter\def\f@size{8}\check@mathfonts
\def\maketag@@@#1{\hbox{\m@th\normalsize\normalfont#1}}%
$$
\frac{\mathrm{d}p_i}{\mathrm{d}t} = -\frac{1}{2}\sum_{j \neq i} \left(p_i\phi_{ij} - p_j\phi_{ij}\right) + \sum_{j \neq i} \frac{D_{ij}}{\lvert e_{ij} \rvert^2} (p_j - p_i) + g_ip_i
$$
\endgroup

{\subsection{Modeling Cell–Cell Interaction~Dynamically}}
For temporally resolved single-cell RNA-seq data, previous modeling approaches have typically been developed in the continuous space of $\mathbb{R}^d$. However, the~data inherently exist within a discrete space. {In addition, incorporating cell–cell communication or interaction into these dynamics are important for constructing accurate spatiotemporal developmental landscapes and for advancing our understanding of complex biological systems~\cite{almet2021landscape,cellchat,jin2025cellchat,commot,almet2024inferring,wada2021cell,topolewski2021information}.} Therefore, an~interesting question is how to construct continuous cellular dynamics from a discrete space of interacting cells, e.g.,~those represented by a graph~\cite{gandrillon2021editorial}. 

\textls[-5]{In ~\cite{jiang2022dynamic}, it proposes GraphFP, a~graph Fokker–Planck equation-based method to model cellular dynamics by explicitly considering cell interactions.  Assume that data can be clustered or annotated into $M$ cell types, GraphFP constructs a cell state transition graph ${G}=(V, E)$, where each vertex in $V$ represents a cell type and each edge $\{i, j\}$ in $E$ means the cell type $i$ can transit to cell type $j$. Unlike other methods considering probability distribution in $\mathbb{R}^d$, GraphFP consider the probability distribution on graph ${G}$. Suppose there are $M$ vertices in graph ${G}$,  consider the probability simplex supported on all vertices of ${G}$}\vspace{-6pt}
$$
\mathcal{P}({G})=\{\boldsymbol{p}(t)=(p_i(t))_{i=1}^{M}\mid \sum_{i=1}^{M}p_i(t)=1, p_i(t)\geq 0\}.
$$
The aim is also to transport the distribution from $\boldsymbol{p}_0$ to $\boldsymbol{p}_1$ on $G$, satisfying least action principles. Similar to the continuous space case, one need to define the Fokker–Planck equation and the Wasserstein distance on graph $G$. First, one can define a free energy $\mathcal{F}:\mathcal{P}(G) \rightarrow \mathbb{R}$, then the Fokker–Planck equation can be defined as follows:
\begingroup
\makeatletter\def\f@size{8}\check@mathfonts
\def\maketag@@@#1{\hbox{\m@th\normalsize\normalfont#1}}%
$$
\frac{\rmd p_i(t)}{\rmd t}=\sum_{j\in \mathcal{N}(i)}\left(\frac{\partial \mathcal{F}(\boldsymbol{p})}{\partial p_j}-\frac{\partial \mathcal{F}(\boldsymbol{p})}{\partial p_i}\right)g_{ij}(\boldsymbol{p}),
$$
\endgroup
where $\mathcal{N}(i)$ is the neighbor set of vertex $i$ and $g_{ij}(\boldsymbol{p})$ satisfy certain constraints that could constructed from $\boldsymbol{p}$~\cite{jiang2022dynamic}. Next, the~discrete L2-Wasserstein distance on graph $G$ between $\boldsymbol{p}_0, \boldsymbol{p}_1 \in \mathcal{P}(G)$ can be defined as\vspace{-6pt}
\begingroup
\makeatletter\def\f@size{8}\check@mathfonts
\def\maketag@@@#1{\hbox{\m@th\normalsize\normalfont#1}}%
$$
\mathcal{W}_{2, G}^2 (\boldsymbol{p}_0, \boldsymbol{p}_1)=\inf_{\mathcal{F}} \frac{1}{2} \int_{0}^{1} \sum_{i,j \in E}\left(\frac{\partial \mathcal{F}(\boldsymbol{p})}{\partial p_j}-\frac{\partial \mathcal{F}(\boldsymbol{p})}{\partial p_i}\right)^2g_{ij}(\boldsymbol{p}).
$$
\endgroup
Next, the~target is to find the minimum energy path. In~~\cite{jiang2022dynamic}, they parameterize $\mathcal{F}$ by a linear energy form\vspace{-12pt}
\begingroup
\makeatletter\def\f@size{8}\check@mathfonts
\def\maketag@@@#1{\hbox{\m@th\normalsize\normalfont#1}}%
$$
\begin{aligned}
\mathcal{F}(\boldsymbol{p} \mid \boldsymbol{\Phi}, \boldsymbol{W}) & =\mathcal{V}(\boldsymbol{p})+\mathcal{W}(\boldsymbol{p})+\beta \mathcal{H}(\boldsymbol{p}), \\
& =\sum_{i=1}^n \Phi_i p_i+\frac{1}{2} \sum_{i=1}^n \sum_{j=1}^n w_{i j} p_i p_j+\beta \sum_{i=1}^n p_i \log p_i, \\
& =\boldsymbol{p}^T \boldsymbol{\Phi}+\frac{1}{2} \boldsymbol{p}^T \boldsymbol{W} \boldsymbol{p}+\beta \sum_{i=1}^n p_i \log p_i,
\end{aligned}
$$
\endgroup
where $\boldsymbol{\Phi}=(\Phi_i)_{i=1}^M$, $\boldsymbol{W}=(w_{i,j})_{1\leq i,j \leq M}$ represents the \textbf{interactions} among cell types, and~$\beta \geq 0$ is a hyper-parameter. After~this parametrization, one denotes the parameters of the free energy as $\boldsymbol{\theta}=\{\boldsymbol{\Phi}, \boldsymbol{W}\}$, and~the goal is to find the parameter $\boldsymbol{\theta}$ such that \vspace{-6pt}
$$
\begin{aligned}
&\boldsymbol{\theta}^*=\arg \min _{\boldsymbol{\theta}} \quad \int_{t_1}^{t_f} \frac{1}{2} \sum_{\{i, j\} \in E}\left(\frac{\partial \mathcal{F}(\boldsymbol{p})}{\partial p_i}-\frac{\partial \mathcal{F}(\boldsymbol{p})}{\partial p_j}\right)^2 \cdot g_{i j}(\boldsymbol{p}(t)) d t,\\
 \end{aligned}
 $$
 subject to the constraints \vspace{-6pt}
 $$
 \begin{aligned}
\frac{d \boldsymbol{p}(t)}{d t} & =\left(\sum_{j \in \mathcal{N}(i)}\left(\frac{\partial \mathcal{F}(\boldsymbol{p})}{\partial p_j}-\frac{\partial \mathcal{F}(\boldsymbol{p})}{\partial p_i}\right) g_{i j}(\boldsymbol{p}(t))\right)_{i=1}^M \\
\boldsymbol{p}(1) & =\boldsymbol{p}_1
\end{aligned}
$$
This problem can be solved by the adjoint method. Once these dynamics are solved, one can then use them for downstream analysis, e.g.,~cell–cell interaction, probability flow of cell types, and~the potential energy~\cite{jiang2022dynamic}. 

Looking ahead, we anticipate that important future directions based on GraphFP include the expansion of the current framework to achieve single-cell resolution instead of cellular types, incorporating the matching of unnormalized distribution results from cell proliferation and death, and~extending the model to spatial~transcriptomics.  

\vspace{1em}

\subsection{Reconstructing  Waddington Developmental~Landscapes}
Waddington’s landscape metaphor is a widely recognized framework to depict the cell fate decision process. This conceptual model suggests that metastable cellular states are analogous to wells within a potential landscape, and~transitions between these states can be understood as movements or “hops” between these potential wells. While the development of such potential landscapes has been extensively explored~\cite{ao2004potential,landscape_PZ, landscape_Shi, epr, lichunhe2013quantifying, Lichunhe2010potential, lichunhe2014landscape, bian2023improved, bian2024quantifying, zhou2024spatial, zhou2024revealing, schiebinger2021reconstructing, torregrosa2021mechanistic}, effectively constructing these landscapes using single-cell omics data remains the major challenge. In~recent works~\cite{scRNA_Wang_landscape,scRNA_Wang_landscape_pnas}, the~authors utilize RNA velocity to construct a vector field from the snapshot scRNA seq data and then compute the potential landscape based on the Boltzmann distribution-like relations proposed by Wang~et~al.~\cite{wang2008potential}. To~be precise, the~landscape is characterized by the expression
$$
U=-\sigma^2 \log p_{\text{ss}}/2
$$
where $p_{\text{ss}}$  represents the steady-state probability density function (PDF) that satisfies the steady-state Fokker–Planck equation:
$
-\nabla \cdot (p_{ss} \boldsymbol{b}) +\frac{\sigma^2}{2} \Delta p_{\text{ss}}= g p_{\text{ss}}
$.
For the temporal scRNA-seq data, following~\cite{DeepRUOT,sflowmatch}, it enables a natural inference of the time-evolving potential energy landscape by leveraging the learned log-density function. Specifically, one can define the landscape at time  $t$  as\vspace{-6pt}
$$
U(\boldsymbol{x}, t) = -\frac{\sigma^2(t)}{2} \log p(\boldsymbol{x}, t).
$$
Regions of lower energy correspond to more stable cell fates, providing a quantitative measure of stability in the cellular state~space.
\subsection{Challenges and Further~Directions}
Integrating multiomics data is critical for comprehensively characterizing cell–cell interaction dynamics and regulatory mechanisms~\cite{stein2021forecasting,cang2024synchronized,demetci2022scot,zhou2023integrating,cao2022unified,SLAT,gao2024graspot,CAST,lahat2015multimodal,liu2023partial}.  Furthermore, aligning both temporal and spatial scales within time-series spatial transcriptomics data (e.g., using non-rigid or non-linear spatial transformations, and~latent space) presents a significant challenge. Additionally, the~application of SDEs in modeling spatial transcriptomics data, and~subsequently constructing spatiotemporal developmental landscapes, represents an important avenue for further~exploration.

\section{Discussion and~Conclusions}\label{sec:discuss}
Inferring dynamical processes from high-throughput single-cell  sequencing data
is a critical problem in understanding cellular development and fate decisions. With~the advancements in sequencing technologies, the~field has evolved from dynamic inference based on snapshot single-cell RNA sequencing (scRNA-seq) data to inferring dynamics from temporally resolved scRNA-seq data. Moreover, the~development of spatial transcriptomics and time-series spatial transcriptomics (ST) data now offers the potential to decode the spatiotemporal developmental trajectories of single cells and construct their spatiotemporal dynamics. In~this review, we have focused on dissecting biological data through the lens of dynamical systems models, specifically investigating how various kinds of models can be applied to study cellular development and fate~decisions.

{When presenting existing approaches, we chose various types of dynamical systems modeling approaches as the main focus, providing a systematic overview of their applications across different contexts. Specifically, we examine the utility and limitations of dynamic modeling techniques in four distinct data scenarios: (1) single time-point scRNA-seq data, (2) multi time-point scRNA-seq data, (3) single time-point spatial transcriptomics data, and~(4) multi time-point spatial transcriptomics data, i.e.,~spatiotemporal single-cell data. For~each data type, we explore how different modeling paradigms—ranging from discrete Markov chain models to continuous Ordinary Differential Equations (ODEs), Stochastic Differential Equations (SDEs), and~Partial Differential Equations (PDEs)—can be used to study the underlying biological processes embedded in the snapshot and high-dimensional nature of sequencing data.}

{For static single snapshot scRNA-seq data, where explicit temporal information is unavailable, top-down discrete-state models such as Markov chains have been widely adopted to infer latent cell state transitions and developmental trajectories. We also address how bottom-up mechanism models like RNA velocity methods provide insights into modeling complex state transition dynamics. When explicit temporal resolution is introduced as in time-series scRNA-seq datasets, continuous dynamical models become more suitable. Incorporating the dynamical optimal transport (OT) theoretical framework, the~Fokker–Planck PDE-based models allow us to track the evolution of cellular states over time.  The~integration of spatial transcriptomics adds another dimension to the modeling challenge. In~the case of multi-time-point spatial transcriptomics, we review emerging approaches that combine dynamical OT and geometric transformation to simultaneously account for trajectory inference and batch correction across time points.}

{Beyond summarizing the mathematical aspects of existing models, we also provide a forward-looking perspective on potential future directions. For~instance, integrating both discrete and continuous models, as~well as cellular interaction effects, could provide new insights to handle more realistic dynamics with increasing biological interpretability. Additionally, the~use of multi-modal omics could also enhance the resolution of dynamical models. To~sum up, by~examining the intersection of dynamical systems theory and single-cell data modeling, this review provides conceptual and methodological insights that may inspire the development of novel algorithms to dissect the spatiotemporal dynamics underlying single-cell sequencing data.}

Due to the limited scope of the current review, several important aspects of the dynamical models of scRNA-seq have not been discussed thoroughly here and remain for further exploration. Firstly, lineage tracing plays a critical role in understanding cellular history and developmental trajectories, and~integrating such data into trajectory inference could provide deeper insights into cellular fate decisions~\cite{wagner2020lineage,weinreb2020lineage,Geofftrajectory}. Secondly, incorporating gene regulatory networks (GRNs)~\cite{GRN1,GRN2,StephenZhang_reference_fitting, stumpf2021inferring, akers2021gene,zhao2024optimal,yang2025topological} into spatiotemporal trajectory inference is an exciting avenue for future research, as~it could enhance the understanding of the regulatory mechanisms driving cellular transitions. Lastly, the~concepts of dynamic network biomarkers (DNBs) and critical transitions~\cite{DNB1,DNB2,DNB3,DNB4} are promising for understanding cellular fate shifts, particularly in disease progression and cellular~development.

Overall, this review demonstrates how dynamic modeling approaches can provide insight into the underlying biological processes underlying single-cell transcriptomics, spatial transcriptomics, and~their temporal extensions. In~the future, these techniques, when combined with machine learning and other computational advancements, will enable more comprehensive models of cellular dynamics, promising new therapeutic strategies and a deeper understanding of development, disease, and~tissue~regeneration.

\vspace{6pt}




\authorcontributions{ Conceptualization, Z.Z., P.Z. and~T.L.; investigation, Z.Z., Y.S., Q.P., P.Z. and~T.L.;   writing---original draft preparation, Z.Z., Y.S., Q.P. and~P.Z.; writing---review and editing,  Z.Z., Y.S., Q.P., P.Z. and~T.L.; visualization, Z.Z., Y.S., Q.P. and~P.Z.; supervision, P.Z. and T.L.; funding acquisition, P.Z. and T.L. All authors have read and agreed to the published version of the manuscript.}

\funding{This work was supported by the National Key R\&D Program of China (No. 2021YFA1003301 to T.L.) and 
 National Natural Science Foundation of China (NSFC No. 12288101 to T.L. \& P.Z., and~8206100646, T2321001 to P.Z.). }






\conflictsofinterest{The authors declare no conflicts of~interest. }

\abbreviations{{Abbreviations}}{{
The following abbreviations are used in this manuscript:}
\\

\noindent 
\begin{tabular}{@{}ll}
{scRNA-seq} & {single-cell RNA sequencing}\\
{SDE}     & {Stochastic Differential Equation}\\
{ODE}     & {Ordinary Differential Equation}\\
{PDE}     & {Partial Differential Equation}\\
{OT}      & {Optimal Transport}\\
{ICA}     & {Independent Component Analysis}\\
{PCA}     & {Principle Component Analysis}\\
{MST}     & {Minimum Spanning Tree}\\
{MCE}     & {Markov Chain Entropy}\\
{GPCCA}   & {Generalized Perron Cluster Cluster Analysis}\\
{EM}      & {Expectation Maximum}\\
{VAE}     & {Variational Autoencoder}\\
\end{tabular}

\noindent
\begin{tabular}{@{}ll} 
{RKHS}    & {Reproducing Kernel Hilbert Space}\\
{CNF}     & {Continuous Normalizing Flow}\\
{FM}      & {Flow Matching}\\
{CFM}     & {Conditional Flow Matching}\\
{SB}      & {Schrödinger Bridge}\\
{RUOT}    & {Regularized Unbalanced Optimal Transport}\\
{GWOT}    & {Gromov–Wasserstein optimal transport}\\
{FGWOT~~~~~}   & {Fused Gromov–Wasserstein optimal transport
}
\end{tabular}
}

\begin{adjustwidth}{-\extralength}{0cm}
\reftitle{References}

\PublishersNote{}
\end{adjustwidth}
\end{document}